\documentclass[aps,prd,twocolumn,superscriptaddress,showpacs]{revtex4-1}

\usepackage{graphicx}% Include figure files
\usepackage{dcolumn}% Align table columns on decimal point
\usepackage{bm}% bold math
\usepackage{lineno}
\usepackage{setspace}
\usepackage{multirow}
\usepackage{slantsc}
\usepackage{longtable}
\usepackage{subfigure}

\newcommand{\bq}{\textit{b}}
\newcommand{\met}{{\not \!\!E}_T}

\begin{document}
\title{
Measurement of the {\boldmath$t\bar t$} production cross section 
in {\boldmath{$p\bar p$}} collisions at {\boldmath$\sqrt{s}=1.96$}~TeV using 
events with large missing transverse energy and jets
}
\affiliation{Institute of Physics, Academia Sinica, Taipei, Taiwan 11529, Republic of China} 
\affiliation{Argonne National Laboratory, Argonne, Illinois 60439, USA} 
\affiliation{University of Athens, 157 71 Athens, Greece} 
\affiliation{Institut de Fisica d'Altes Energies, Universitat Autonoma de Barcelona, E-08193, Bellaterra (Barcelona), Spain} 
\affiliation{Baylor University, Waco, Texas 76798, USA} 
\affiliation{Istituto Nazionale di Fisica Nucleare Bologna, $^z$University of Bologna, I-40127 Bologna, Italy} 
\affiliation{University of California, Davis, Davis, California 95616, USA} 
\affiliation{University of California, Los Angeles, Los Angeles, California 90024, USA} 
\affiliation{Instituto de Fisica de Cantabria, CSIC-University of Cantabria, 39005 Santander, Spain} 
\affiliation{Carnegie Mellon University, Pittsburgh, Pennsylvania 15213, USA} 
\affiliation{Enrico Fermi Institute, University of Chicago, Chicago, Illinois 60637, USA}
\affiliation{Comenius University, 842 48 Bratislava, Slovakia; Institute of Experimental Physics, 040 01 Kosice, Slovakia} 
\affiliation{Joint Institute for Nuclear Research, RU-141980 Dubna, Russia} 
\affiliation{Duke University, Durham, North Carolina 27708, USA} 
\affiliation{Fermi National Accelerator Laboratory, Batavia, Illinois 60510, USA} 
\affiliation{University of Florida, Gainesville, Florida 32611, USA} 
\affiliation{Laboratori Nazionali di Frascati, Istituto Nazionale di Fisica Nucleare, I-00044 Frascati, Italy} 
\affiliation{University of Geneva, CH-1211 Geneva 4, Switzerland} 
\affiliation{Glasgow University, Glasgow G12 8QQ, United Kingdom} 
\affiliation{Harvard University, Cambridge, Massachusetts 02138, USA} 
\affiliation{Division of High Energy Physics, Department of Physics, University of Helsinki and Helsinki Institute of Physics, FIN-00014, Helsinki, Finland} 
\affiliation{University of Illinois, Urbana, Illinois 61801, USA} 
\affiliation{The Johns Hopkins University, Baltimore, Maryland 21218, USA} 
\affiliation{Institut f\"{u}r Experimentelle Kernphysik, Karlsruhe Institute of Technology, D-76131 Karlsruhe, Germany} 
\affiliation{Center for High Energy Physics: Kyungpook National University, Daegu 702-701, Korea; Seoul National University, Seoul 151-742, Korea; Sungkyunkwan University, Suwon 440-746, Korea; Korea Institute of Science and Technology Information, Daejeon 305-806, Korea; Chonnam National University, Gwangju 500-757, Korea; Chonbuk National University, Jeonju 561-756, Korea} 
\affiliation{Ernest Orlando Lawrence Berkeley National Laboratory, Berkeley, California 94720, USA} 
\affiliation{University of Liverpool, Liverpool L69 7ZE, United Kingdom} 
\affiliation{University College London, London WC1E 6BT, United Kingdom} 
\affiliation{Centro de Investigaciones Energeticas Medioambientales y Tecnologicas, E-28040 Madrid, Spain} 
\affiliation{Massachusetts Institute of Technology, Cambridge, Massachusetts 02139, USA} 
\affiliation{Institute of Particle Physics: McGill University, Montr\'{e}al, Qu\'{e}bec, Canada H3A~2T8; Simon Fraser University, Burnaby, British Columbia, Canada V5A~1S6; University of Toronto, Toronto, Ontario, Canada M5S~1A7; and TRIUMF, Vancouver, British Columbia, Canada V6T~2A3} 
\affiliation{University of Michigan, Ann Arbor, Michigan 48109, USA} 
\affiliation{Michigan State University, East Lansing, Michigan 48824, USA}
\affiliation{Institution for Theoretical and Experimental Physics, ITEP, Moscow 117259, Russia}
\affiliation{University of New Mexico, Albuquerque, New Mexico 87131, USA} 
\affiliation{Northwestern University, Evanston, Illinois 60208, USA} 
\affiliation{The Ohio State University, Columbus, Ohio 43210, USA} 
\affiliation{Okayama University, Okayama 700-8530, Japan} 
\affiliation{Osaka City University, Osaka 588, Japan} 
\affiliation{University of Oxford, Oxford OX1 3RH, United Kingdom} 
\affiliation{Istituto Nazionale di Fisica Nucleare, Sezione di Padova-Trento, $^{aa}$University of Padova, I-35131 Padova, Italy} 
\affiliation{LPNHE, Universite Pierre et Marie Curie/IN2P3-CNRS, UMR7585, Paris, F-75252 France} 
\affiliation{University of Pennsylvania, Philadelphia, Pennsylvania 19104, USA}
\affiliation{Istituto Nazionale di Fisica Nucleare Pisa, $^{bb}$University of Pisa, $^{cc}$University of Siena and $^{dd}$Scuola Normale Superiore, I-56127 Pisa, Italy} 
\affiliation{University of Pittsburgh, Pittsburgh, Pennsylvania 15260, USA} 
\affiliation{Purdue University, West Lafayette, Indiana 47907, USA} 
\affiliation{University of Rochester, Rochester, New York 14627, USA} 
\affiliation{The Rockefeller University, New York, New York 10065, USA} 
\affiliation{Istituto Nazionale di Fisica Nucleare, Sezione di Roma 1, $^{ee}$Sapienza Universit\`{a} di Roma, I-00185 Roma, Italy} 

\affiliation{Rutgers University, Piscataway, New Jersey 08855, USA} 
\affiliation{Texas A\&M University, College Station, Texas 77843, USA} 
\affiliation{Istituto Nazionale di Fisica Nucleare Trieste/Udine, I-34100 Trieste, $^{ff}$University of Trieste/Udine, I-33100 Udine, Italy} 
\affiliation{University of Tsukuba, Tsukuba, Ibaraki 305, Japan} 
\affiliation{Tufts University, Medford, Massachusetts 02155, USA} 
\affiliation{University of Virginia, Charlottesville, VA  22906, USA}
\affiliation{Waseda University, Tokyo 169, Japan} 
\affiliation{Wayne State University, Detroit, Michigan 48201, USA} 
\affiliation{University of Wisconsin, Madison, Wisconsin 53706, USA} 
\affiliation{Yale University, New Haven, Connecticut 06520, USA} 
\author{T.~Aaltonen}
\affiliation{Division of High Energy Physics, Department of Physics, University of Helsinki and Helsinki Institute of Physics, FIN-00014, Helsinki, Finland}
\author{B.~\'{A}lvarez~Gonz\'{a}lez$^v$}
\affiliation{Instituto de Fisica de Cantabria, CSIC-University of Cantabria, 39005 Santander, Spain}
\author{S.~Amerio}
\affiliation{Istituto Nazionale di Fisica Nucleare, Sezione di Padova-Trento, $^{aa}$University of Padova, I-35131 Padova, Italy} 

\author{D.~Amidei}
\affiliation{University of Michigan, Ann Arbor, Michigan 48109, USA}
\author{A.~Anastassov}
\affiliation{Northwestern University, Evanston, Illinois 60208, USA}
\author{A.~Annovi}
\affiliation{Laboratori Nazionali di Frascati, Istituto Nazionale di Fisica Nucleare, I-00044 Frascati, Italy}
\author{J.~Antos}
\affiliation{Comenius University, 842 48 Bratislava, Slovakia; Institute of Experimental Physics, 040 01 Kosice, Slovakia}
\author{G.~Apollinari}
\affiliation{Fermi National Accelerator Laboratory, Batavia, Illinois 60510, USA}
\author{J.A.~Appel}
\affiliation{Fermi National Accelerator Laboratory, Batavia, Illinois 60510, USA}
\author{A.~Apresyan}
\affiliation{Purdue University, West Lafayette, Indiana 47907, USA}
% ...
\collaboration{CDF Collaboration\footnote{With visitors from $^a$University of Massachusetts Amherst, Amherst, Massachusetts 01003,
$^b$Istituto Nazionale di Fisica Nucleare, Sezione di Cagliari, 09042 Monserrato (Cagliari), Italy,
$^c$University of California Irvine, Irvine, CA  92697, 
$^d$University of California Santa Barbara, Santa Barbara, CA 93106
$^e$University of California Santa Cruz, Santa Cruz, CA  95064,
$^f$CERN,CH-1211 Geneva, Switzerland,
$^g$Cornell University, Ithaca, NY  14853, 
$^h$University of Cyprus, Nicosia CY-1678, Cyprus, 
$^i$University College Dublin, Dublin 4, Ireland,
$^j$University of Fukui, Fukui City, Fukui Prefecture, Japan 910-0017,
$^k$Universidad Iberoamericana, Mexico D.F., Mexico,
$^l$Iowa State University, Ames, IA  50011,
$^m$University of Iowa, Iowa City, IA  52242,
$^n$Kinki University, Higashi-Osaka City, Japan 577-8502,
$^o$Kansas State University, Manhattan, KS 66506,
$^p$University of Manchester, Manchester M13 9PL, England,
$^q$Queen Mary, University of London, London, E1 4NS, England,
$^r$Muons, Inc., Batavia, IL 60510,
$^s$Nagasaki Institute of Applied Science, Nagasaki, Japan, 
$^t$National Research Nuclear University, Moscow, Russia,
$^u$University of Notre Dame, Notre Dame, IN 46556,
$^v$Universidad de Oviedo, E-33007 Oviedo, Spain, 
$^w$Texas Tech University, Lubbock, TX  79609, 
$^x$Universidad Tecnica Federico Santa Maria, 110v Valparaiso, Chile,
$^y$Yarmouk University, Irbid 211-63, Jordan,
$^{gg}$On leave from J.~Stefan Institute, Ljubljana, Slovenia, 
}}
\noaffiliation

\author{The CDF Collaboration}

\vspace*{2.0cm}

\begin{abstract}
In this paper we report a measurement of the 
$t\bar t$ production cross section 
in $p\bar p$ collisions at $\sqrt{s}=1.96$~TeV
using data corresponding to an integrated luminosity of 2.2 
fb$^{-1}$ collected with the CDF II detector at the 
Tevatron accelerator.
We select events with significant
missing transverse energy and high jet multiplicity. 
This measurement vetoes the presence 
of explicitly identified electrons and muons, thus enhancing the tau contribution of 
$t \bar t$ decays. Signal events are discriminated from 
the background using a neural network, and 
heavy flavor jets are identified 
by a secondary-vertex tagging algorithm.
We measure a $t\bar t$ production cross section of
$7.99~\pm 0.55 \mbox{ (stat) }\pm 0.76 \mbox{
(syst) }\pm 0.46 \mbox{ (lumi)}$~pb,
assuming a top mass $m_{top} = 172.5$~GeV/$c^2$,
in agreement with previous measurements and 
standard model predictions.
\end{abstract}

\maketitle
\newpage

\section{Introduction}\label{sec:Intro}
In $p\bar p$ collisions at $\sqrt{s} = 1.96$~TeV at the Tevatron,
top quarks are produced mainly in pairs through
quark-antiquark annihilation
and gluon-gluon fusion processes.
In the standard model~(SM), the calculated cross section
for top pair production at the Tevatron center-of-mass 
energy is $7.46^{+0.66}_{-0.80}$~pb~\cite{top_th}
for a top mass of $172.5$~GeV/$c^2$. 
This value 
can be enhanced by new processes beyond the SM such as top
pair production via new massive resonances~\cite{massive_resonance},
while the comparison of the top pair production 
cross section in different decay channels
can be sensitive to the presence of  top decays 
via a charged Higgs boson~\cite{charged_higgs}.
Thus, a precise measurement of the top pair production cross section
is an important test of the SM. Both CDF and D0 have performed 
many measurements of this quantity
in different $t \bar t$ final states: the most recent published 
results, both measured in the decay channel with leptons and 
jets assuming $m_{top} = 172.5$~GeV/$c^2$, are 7.70 $\pm$ 0.52 pb for 
CDF~\cite{cdf_recent_result}  and 
7.78~$^{+0.77}_{-0.64}$ pb for D0~\cite{d0_recent_result}.

As the Cabibbo-Kobayashi-Maskawa matrix element $V_{tb}$
is close to unity ~\cite{d0_Vtb, cdf_Vtb} and
the top mass $m_{top}$ is larger
than the sum of the $W$ boson and bottom quark~($b$) masses,
in the SM the $t\to W b$ decay is dominant
and has a branching ratio of about $100\%$.
Since the $W$ subsequently decays either to
a quark-antiquark pair or to a lepton-neutrino pair,
the resulting
top pair production
final states can be classified by the number of
energetic charged leptons and the number of jets.
When only one $W$ decays leptonically, 
the $t\bar  t$ event is characterized by
the presence of
a charged lepton, missing energy 
due to the undetected neutrino,
and four high 
energy jets, two of which originate from $\bq$ quarks.
In this {\em lepton plus jets} channel, one selects events with
an energetic electron or muon.  For this paper we focus 
on an inclusive high-momentum neutrino 
signature of large missing energy accompanied by jets. 
By not explicitly requiring leptons, 
our measurement is sensitive to all $W$ 
leptonic decay modes including $\tau$ decays of $W$'s: about 40\% of 
the signal sample obtained after the kinematic selection contains events with 
a $\tau$ lepton in the final state.
To ensure our measurement is statistically 
independent from other CDF results~\cite{prevCDFMeas}, 
we veto events with high-momentum electrons or muons 
as well as multijet events with no leptons
(\textit{all-hadronic} $t\bar t$ decays).
This choice is expected to improve the final
CDF combined cross section value: 
the  previous 311 pb$^{-1}$ result~\cite{cdf_metJetCortiana}
carried a weight of about 17\% in the combination~\cite{cdf_xsec_combo}.

One of the major challenges of this measurement 
is  the large background from
QCD multijet processes
and
electroweak production
of $W$ bosons associated with $\bq$ and $c$ jets (heavy flavor jets),
which dominates the signal by two orders of magnitude before any selection.
In order to improve the signal to background ratio (S/B),
a neural network is trained to identify
the kinematic and topological
characteristics of SM $t\bar t$ events,
and is applied to data to select
a signal-rich sample.
Top quarks are then identified by
their distinctive decay into $b$ quarks. Jets originating from
$b$ quarks ($b$ jets) are selected~(``tagged'') 
by their displaced vertex as defined by the {\scshape{secvtx}} algorithm~\cite{secvtx}.
After evaluating 
the average number of $\bq$-tagged jets for
$t\bar t$ events using a Monte Carlo~(MC) simulation,
the number of signal events in the sample and the
corresponding cross section are measured by counting the number of 
$\bq$-tagged jets
in the sample selected by the neural network.
The number of background $\bq$-tagged jets is estimated
using per-jet  parametrized probabilities of $\bq$-jet
identification, measured directly from data,
rather than relying on
theoretical prediction of cross sections and
MC simulations. The results reported here are based on data taken
between March 2002 and August 2007, corresponding to an
integrated luminosity of $2.2$~fb$^{-1}$,
recorded by the CDF experiment at Fermilab.

The organization of the paper is the following:
Sec.~\ref{sec:CDFdetector}  contains a brief description of 
the CDF II detector and of
the trigger requirements used for this analysis.
The preliminary clean-up cuts applied to data are
 described in Sec.~\ref{sec:prereq}, followed by
the discussion of the data-driven background parametrization 
in Sec.~\ref{sec:Bgnd}.
The kinematic variables characterizing the  
missing energy plus jets final state 
 and the neural network based sample selection 
 are described in Sec.~\ref{sec:SignalCharact} and
in Sec.~\ref{sec:SignalSel} respectively.
We conclude the description of this measurement with 
a summary of the different sources of systematic uncertainties 
in Sec.~\ref{sec:Syst}, while the cross section measurement is presented 
in Sec.~\ref{sec:XsecMeas}.

\section{The CDF Detector and Trigger System}\label{sec:CDFdetector}
CDF\,II is a general--purpose, azimuthally and forward-backward symmetric 
detector located at the Tevatron $p \bar p$ collider at Fermilab. 
It consists of a charged-particle tracking system immersed 
in a 1.4~T magnetic field. The solenoid is surrounded by 
calorimeters and muon detectors~\cite{CDF}. 
The CDF\,II coordinate system uses $\theta$ and $\phi$ as the polar and
azimuthal angles respectively, defined with respect to the proton beam axis
direction,~$z$. The $x$-axis points towards the center of the accelerator 
while the $y$ axis upwards from the beam. The pseudorapidity $\eta$ is defined 
as $\eta\equiv-\ln[\tan(\theta /2)]$. 
The transverse momentum of a particle is $p_T = p \sin\theta$
 and its transverse energy $E_T = E \sin\theta$.
The missing transverse energy $\met$ measures the transverse energy of the 
neutrinos via the imbalance of the energy detected in the calorimeters; 
it is defined by $\met$ = $|{\not\!\! \vec{E}_T}|$ where
${\not\!\! \vec{E}_T} = - \sum_{i} E_T^i \hat{n}_i$ , the index $i$ runs over 
the calorimeter tower number 
and $\hat{n}_i$ is a unit vector perpendicular to the beam axis and 
pointing at the \textit{i}-th calorimeter tower. 
The tracking system is composed of 8 layers of silicon microstrip detectors,
extending from 1.6 cm to 28 cm and covering up to $|\eta| <$ 2.0, 
surrounded by a 3.1 m long open cell drift chamber, 
providing $|\eta|$ coverage up to 1.0.
Using information from the silicon detectors, the primary interaction vertex
is reconstructed with a precision of $\sim$~15~$\mu$m 
in the plane transverse to the beam~\cite{vertex}.
The energy of the particles traversing the detector is measured by 
electromagnetic and hadronic calorimeters segmented into projective 
towers covering up to $|\eta| < $ 3.6. In the central region  ($|\eta| <$ 1.1) 
the calorimetric towers are 15$^{\circ}$ wide in $\phi$ and 0.1 in $\eta$; 
in the forward region  
(1.1 $< |\eta| < $ 3.6) the towers are 7.5$^{\circ}$ wide in azimuthal angle 
for $|\eta| < $ 2.1 and 15$^{\circ}$ for $|\eta| > $ 2.1. 
The electromagnetic section is made of lead-scintillator plates, while 
the hadronic section uses iron-scintillator ones. The transverse 
profile of electromagnetic 
showers is measured by  
proportional chambers and scintillating strip 
detectors.
Muons are detected up to $|\eta| < $ 0.6 by drift chambers located 
outside the hadronic calorimeters, behind a 60 cm iron shield. 
Additional drift chambers and scintillator detectors provide muon 
detection up to $|\eta| < $ 1.5.

The CDF\,II trigger system~\cite{CDFtrigger} has a three 
level architecture designed to operate at 2.53~MHz
and reduce the data rate to approximately 120~Hz 
to be written on tape.
The data used in this measurement are collected with a purely calorimetric
 trigger, described in Sec.~\ref{sec:multiJetTrig}.
 At Level-1 (L1) calorimetric towers are merged in pairs 
along $\eta$ to  define \textit{trigger towers}: the L1 can take a decision
 on the energy contribution of individual 
trigger towers, on the sum of the energy of all the towers 
or on the missing transverse energy of the event.
At Level-2 (L2) trigger towers are merged into clusters by a simple clustering 
algorithm~\cite{CDF-TDR},
while at Level-3 (L3) jets of particles are reconstructed by a fixed cone-based 
algorithm~\cite{jetclu}, the radius of the cone in 
the $\eta - \phi$ plane ($\Delta R = \sqrt{\Delta \eta^2 + \Delta \phi^2}$)
 being 0.4, 0.7 or 1 depending  on the specific trigger.

\subsection{The Multijet trigger}\label{sec:multiJetTrig}
The data used in this analysis are collected by a 
multijet trigger. 
This trigger requires at L1 the presence of at least one central 
trigger tower with $E_T \ge $~10~GeV, and
at L2 at least four calorimetric clusters with $E_T \ge$ 15~GeV each
and a total transverse energy greater than 175 GeV. The latter threshold was 
125~GeV before February 2005 and was increased to reduce the 
trigger rate at higher instantaneous luminosity.  
Finally, at L3,  
at least four jets with $E_T \ge $10~GeV ($\Delta R$ = 0.4) are required. 
This trigger was  specifically designed to 
collect all-hadronic 
$t\bar t$ events, where the final state nominally consists of six
jets, but has a large acceptance also on 
final states characterized by $\met$ and 
high jet multiplicity. Moreover the collected sample is uncorrelated with 
those used for top cross section measurements in the 
lepton plus jets final state, 
which are selected by requiring the presence of a high
 momentum lepton. 
The choice of this trigger is also driven
 by the analysis strategy: 
the top cross section measurement is  
performed by counting the number of $\bq$-tagged jets from
top decay in the final sample and 
the multijet trigger provides a sample that is
unbiased with respect to the $b$-tagging algorithm, as 
it does not apply any requirement on tracks.

\section{Preliminary Requirements}\label{sec:prereq}
Events satisfying the trigger requirements are used in the analysis only
if they were collected with fully operational tracking detectors, calorimeters 
and muon systems,
and if their primary vertex is located within $\pm $ 60 cm along $z$ 
from the center of CDF~II detector.
Jets and $\met$ are corrected~\cite{jet_corr} for 
multiple $p \bar p$ interactions 
in the event, non uniformities in the calorimeter response along 
$\eta$, and any non-linearity and energy 
loss in the uninstrumented regions of the calorimeters. 
The $\met$ is also corrected for the 
presence of high-$p_T$ muons.
We consider jets with $\Delta R$ = 0.4, 
corrected transverse energy $E_T\ge 15$~GeV
and $|\eta|\le 2.0$; 
the total transverse energy of the event $\sum E_T$ 
is defined as the sum of all jets $E_T$.

Events are required to have at least three 
jets and no central    
high-$p_T$  ($p_T > $ 20 GeV/c) reconstructed electrons or muons
 to avoid overlaps with other top cross section measurements in lepton and 
jets final states~\cite{lepton_jets}.
In the same way,  overlaps with top all-hadronic 
analyses are avoided by rejecting events with low 
$\met$ significance $\met^{sig}$, 
defined as $\met^{sig} = \met / \sqrt{\sum E_T}$
where $\met$ and $\sum E_T$ are measured in GeV, 
as the resolution on the $\met$ is observed to degrade as a 
function of the total transverse energy of the event.
The $\met^{sig}$ is typically low when $\met$ arises from mismeasurements,
so events are required to have ${\met}^{sig} \ge 3$~GeV$^{1/2}$.

Throughout the paper, the impact of analysis requirements
on signal is evaluated using
inclusive $t \bar t$ samples 
generated with {\sc{pythia}} version~v6.216~\cite{pythia}
and processed through CDF~II detector and full trigger
simulation~\cite{cdfsim}, assuming a 
top mass $m_{top} = 172.5$~GeV/$c^2$ (the most 
recent CDF and D0 combined result on top mass is 
$m_{top} = 173.3 \pm 1.1$~GeV/$c^2$~\cite{topMass}).
After the preliminary cuts, there are 
$94~217$ events remaining in the sample with at least four jets,
with an expected signal to background ratio $S/B$
of $1.4\%$ and
$44~310$ events  in the sample with exactly three jets,
with an expected $S/B$ lower than $0.1\%$. % is $0.07\%$
The latter sample, with a very low $t\bar t$ signal 
content, will be used
to derive the background parametrization in Sec.~\ref{sec:Bgnd}.

\section{Background Parametrization}\label{sec:Bgnd}
In this analysis, top quarks are identified
by their decays into $\bq$ quarks.
However, many processes other than the decay of the top quark 
can give rise to $b$ jets, therefore a procedure to
estimate the number of $\bq$-tagged jets yielded 
by background processes is needed.

\begin{figure*}[]
 \centering
\subfigure[][]{ \label{subfig:tagrate_et}\includegraphics[width=0.3\linewidth]{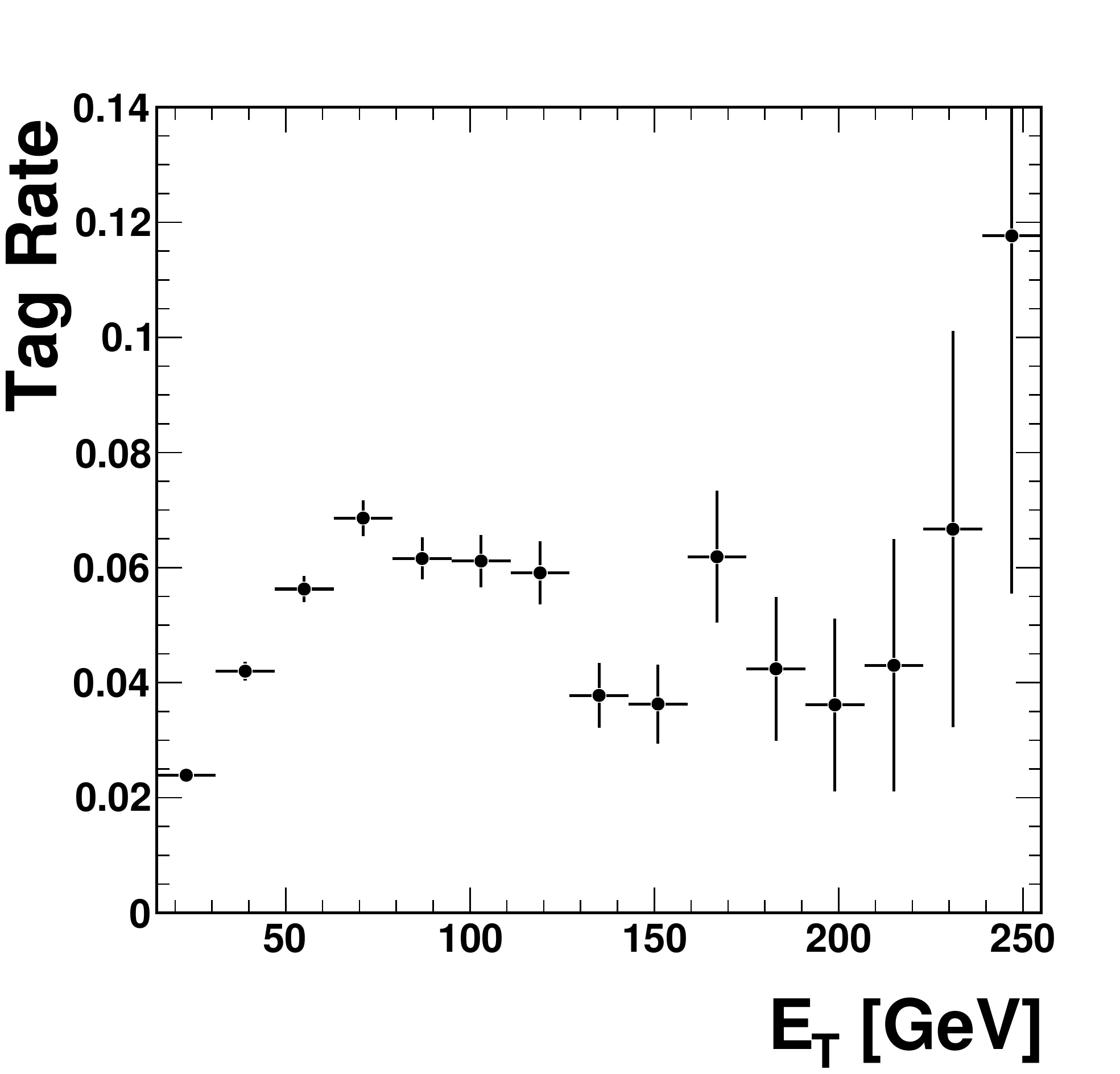}}
\subfigure[][]{ \label{subfig:tagrate_ntrks}\includegraphics[width=.3\linewidth]{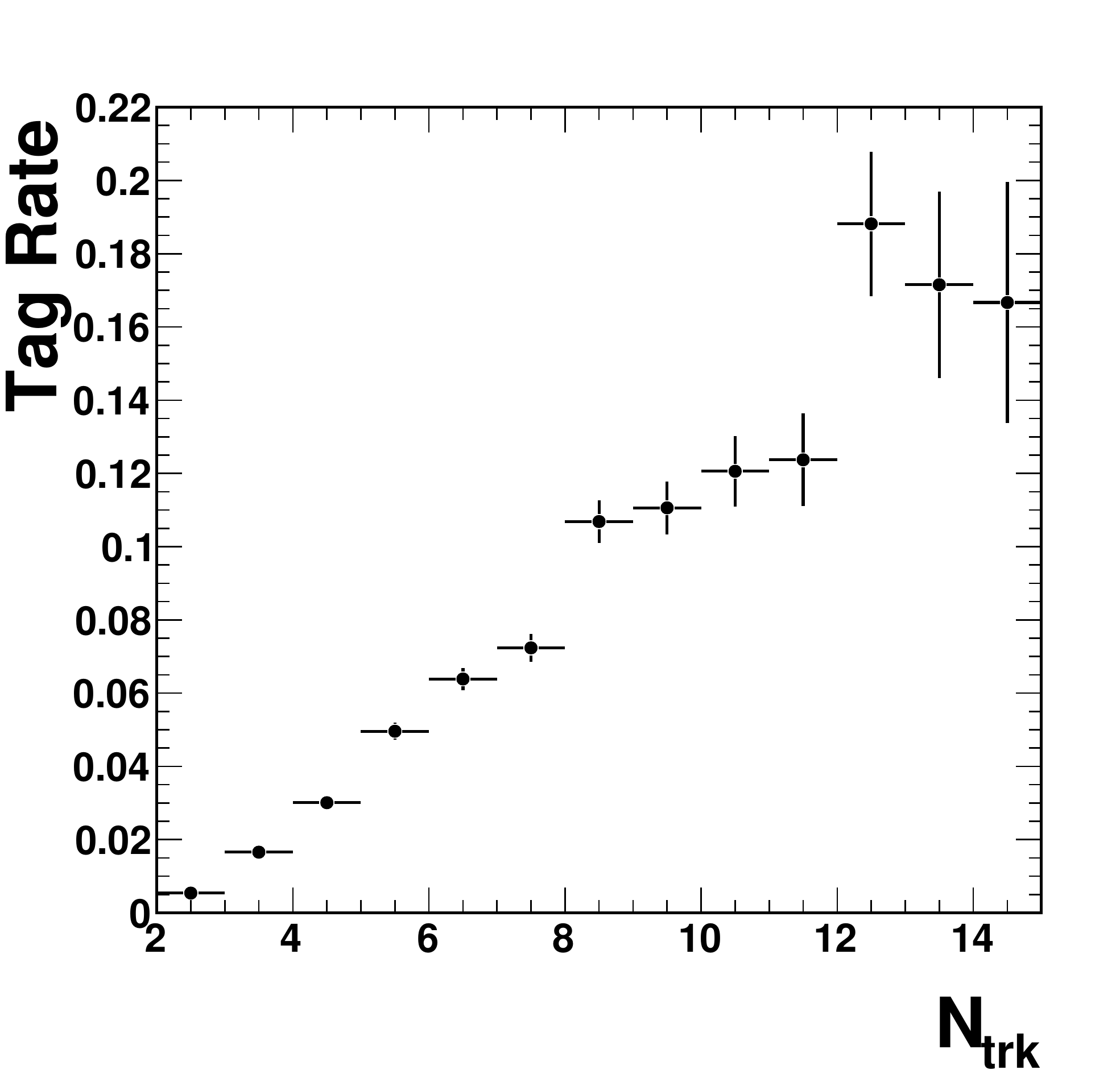}}
\subfigure[][]{ \label{subfig:tagrate_metprj}\includegraphics[width=.3\linewidth]{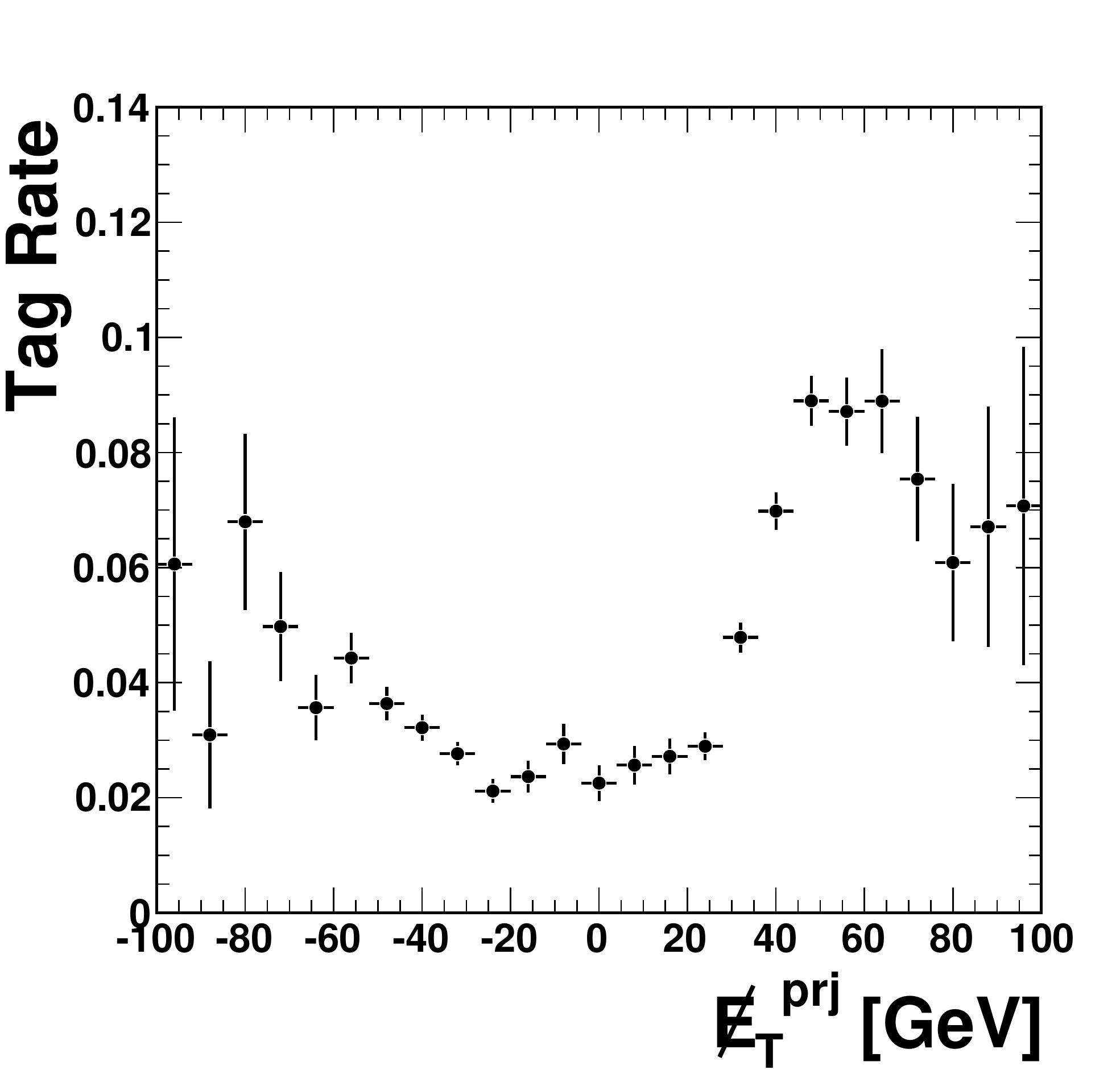}}
 \caption{$b$ tagging rates as a function of jet $E_T$~\subref{subfig:tagrate_et}, 
 	$N_{trk}$~\subref{subfig:tagrate_ntrks} 
   and ${\met}^{prj}$~\subref{subfig:tagrate_metprj} for 
        the data sample with
 	exactly three jets in the event. 
          }
 \label{fig:tagrate1}
 \end{figure*}

The background-prediction method used in this analysis rests 
on the assumption that the probabilities of tagging a $b$ jet in
 $t\bar t$ signal and background processes 
are different: the differences are due 
 to the distinctive properties of $b$ jets produced by the top quark 
 decays, compared to $b$ jets arising from QCD and $W$
 boson plus heavy flavor production processes. 
In this hypothesis, parametrizing the $b$-tag rates as a function 
of jet variables in events depleted of signal 
allows one to predict the number of $b$-tagged jets from 
background processes in any given sample.
The parametrization is derived from a sample of
pure background: the data sample after 
 the prerequisites requirement with exactly three jets,
 where the $t\bar t$ signal fraction is lower than $0.1\%$.
In this sample the per-jet $b$-tagging probability 
is found to depend mainly on jet
$E_T$, the number of good quality tracks contained
in the jet cone $N_{trk}$, and the $\met$ projection along the jet direction
${\met}^{prj}$, defined as ${\met}^{prj} = \met \cos \Delta\phi(\met, jet)$.
These variables are chosen for the parametrization and their
corresponding tag rate dependence is shown in Fig.~\ref{fig:tagrate1}.
We expect the tagging probability to depend on jet $E_T$ and $N_{trk}$  
due to the implementation details of 
the $b$-tagging algorithm~\cite{secvtx}. In more detail, the 
$b$-tagging probability decreases at high jet 
$E_T$ due to the declining yield of tracks 
passing the quality cuts required by the $b$-tagging algorithm;
it increases if a greater number of good quality tracks is associated
to the jet cone. 
The $\met$ projection along the jet direction is correlated 
with the heavy flavor component of the sample and the geometric
properties of the event. Neutrinos from the $\bq$-quark 
semi-leptonic decays force $\met$ to be aligned with the jet 
direction, while neutrinos from $W$ boson decays are more 
likely to be away from jets. For this reason the  
$b$-tag rate is enhanced at high positive values of 
${\met}^{prj}$.
 
 A 3-dimensional $b$-tagging matrix $\mathcal{P}$ is defined 
 using the per-jet $b$-tagging probabilities,
 and its binning is defined to avoid any entries 
with empty denominators.
The matrix assigns the probability that a jet is $\bq$-tagged given its 
$E_T$, $N_{trk}$ and ${\met}^{prj}$.\\
 The total number of expected background 
 $b$-tagged jets $N_{exp}$ in a given 
 data sample can be calculated by summing the
 $b$-tagging probabilities
 over all jets in the selected events:
 \begin{equation}
 N_{exp} = \sum_{i=1}^{N_{events}} \sum_{k=1}^{N_{jets,i}} \mathcal{P}_i (E_T^k, N_{trk}^k, {\met}^{prj,k})
 \label{eq:number_of_tags}
 \end{equation}
 where the index $k$ runs over the number of jets $N_{jets}$
 in the $i$-th event and the index $i$
 runs over all the $N_{events}$ events in the sample.
 Due to the method we used, 
 $N_{exp}$ is obtained
 under the same assumption that 
 allowed the parametrization of the tagging rate,
 i.e. a negligible $t\bar t$ signal content.
For jet multiplicities greater than three,  $N_{exp}$ will 
overestimate the number of $b$-tagged jets from background events, $N_{bkg}$, due 
to the presence of $t \bar t$ events in the sample.
 If $N_{obs}$ is the number of $b$-tagged jets observed in the data
sample and we assume that the difference between 
 $N_{obs}$ and $N_{bkg}$ is due to the presence
 of $t\bar t$ signal, the number of top events in the sample 
 $n_{top}$
 can be evaluated as
 $n_{top} = (N_{obs} - N_{bkg})/\epsilon_{tag}^{ave}$,
 where $\epsilon_{tag}^{ave}$ is the average number of $b$-tagged jets
 per top event, 
 defined as the ratio between the number of $b$-tagged jets
 and the number of $t\bar t$ events and calculated from MC simulation.
 If $n_{evt}$ is the number of events in the sample,
 we can estimate the number of $b$-tagged jets due to the 
background $N_{bkg}$, by rescaling the value $N_{exp}$ predicted by the 
$b$-tagging matrix as follows:
 $N_{bkg} = N_{exp} (n_{evt} - n_{top}) / n_{evt}$.
 Putting the expressions for $N_{bkg}$ and $n_{top}$ together,
 we can evaluate $N_{bkg}$ as the limit of the following iterative formula:
 \begin{equation}
 N_{bkg}  =  \lim_{i\to \infty} N_{exp, i} 
 \end{equation}
with 
\begin{eqnarray}
%         \mbox{ with }  \\
  N_{exp, i}  &=&   N_{exp, 0} \frac{n_{evt} - n_{top}}{n_{evt}} \nonumber \\ 
            &=&   N_{exp, 0} \frac{n_{evt}- 
 		 (N_{obs} - N_{exp, i-1})/\epsilon_{tag}^{ave}
 		 }{n_{evt}}
\end{eqnarray}

 where $N_{exp, 0} = N_{exp}$ is the number - fixed during the iteration - 
 of expected $b$-tagged jets coming from the tag rate parametrization 
 before any correction. In our calculations 
 the iterative procedure stops when
 $ \frac{|N_{exp, i} - N_{exp, i-1}|}{N_{exp, i}} \le 1\%$.
 This correction procedure
 is used in all samples with more than three jets 
 to remove the $t\bar t$ contribution 
 from the background estimation.
 This method does not require knowledge of the top production cross section,
 once we assume that the dependence of
 $\epsilon_{tag}^{ave}$ from the top quark mass is negligible.

 \begin{figure}
 \centering
 \includegraphics[width=0.9\linewidth]{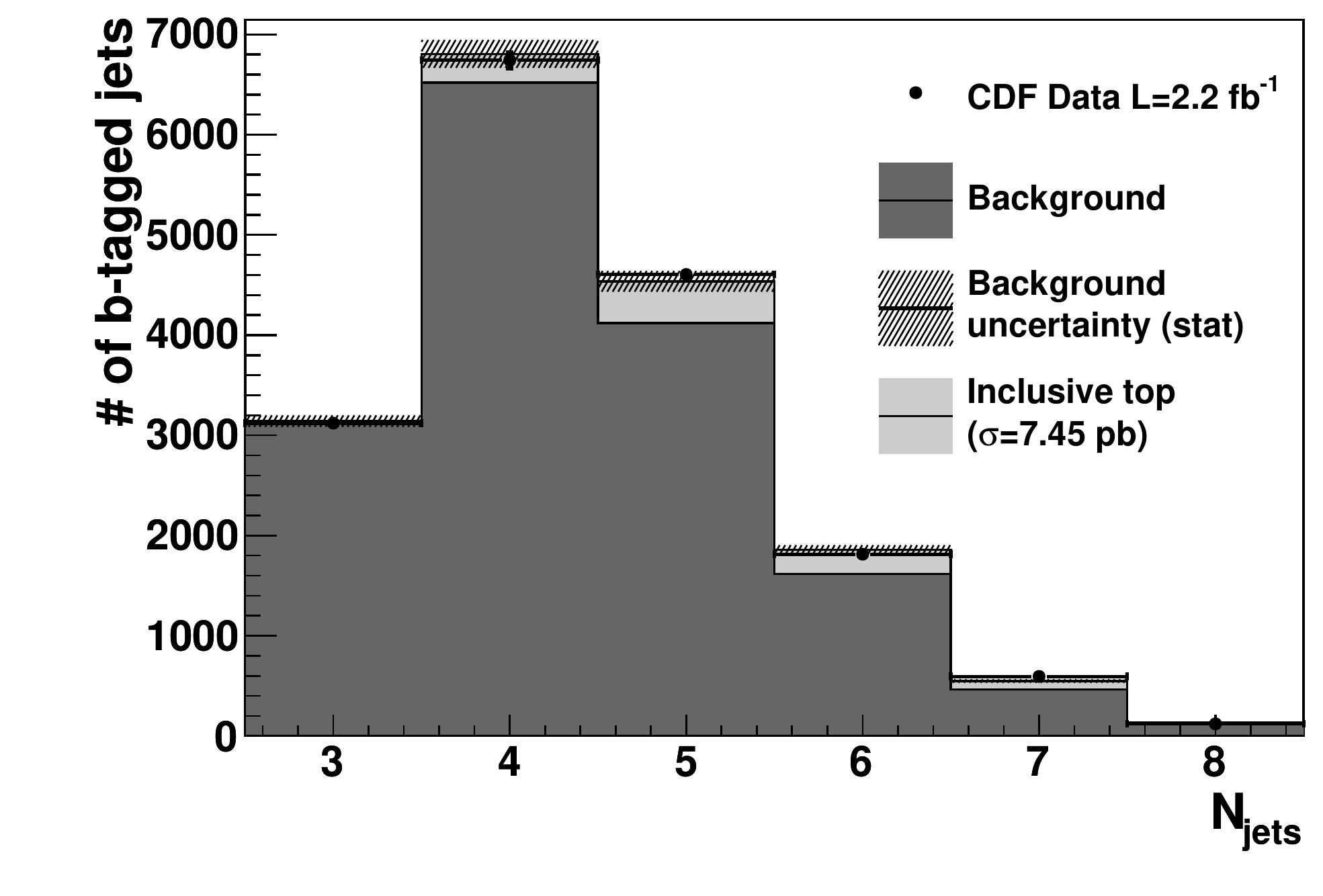}
 \caption{Tagging matrix check 
 	on data after preliminary requirements and	
 	before any additional kinematic selection.
 	Observed and predicted number of $b$-tagged jets as a 
 	function of the jet multiplicity are shown in the figure,
 	statistical errors only.
 	The expected contribution coming from $t\bar t$ events 
	based on the theoretical cross section of $\sigma_{t\bar t}=7.45$~pb
	is also shown.
	Background error bands are centered on the 
    inclusive top plus background prediction.}
 \label{fig:tags_njets}
 \end{figure}
 
 To ensure the correct behavior of the $b$-tagging parametrization, 
 an important check 
 consists in calculating the predicted number of $b$-tagged jets
 in data samples with jet multiplicities higher than three,
 and compare it
 with the actual observed number of $b$-tagged jets.
 The result of this check is shown in Fig.~\ref{fig:tags_njets}.
Taking into account the 
expected contribution to the number of
observed $b$-tagged jets due to the presence of $t\bar t$
events in the sample ($m_{top} = 172.5$~GeV/$c^2$),
 the agreement between the amount of observed and predicted $b$-tagged jets is 
 good in all the jet multiplicity bins, being exactly the same by definition 
 for 3-jet events, on which the parametrization is calculated. 
 
\section{Signal and background characterization}\label{sec:SignalCharact}
In the data sample under investigation, $t\bar t$ pairs 
are overwhelmed by multijet QCD and $W$ + jets events. 
The main feature of the  $t\bar t$ decay channel analyzed here is  
a considerable amount of
$\met$, the only observable signature
of the presence of neutrinos 
from $W$ leptonic decays.
However, missing transverse energy can be also produced 
by jet energy mismeasurements, by semi-leptonic 
decays of $b$ quarks in QCD events and by the leptonic decay of 
the $W$ in $W$ + jets events. 
To discriminate against the possible sources of missing transverse energy on 
a geometric basis we use the quantity 
$\Delta\phi_{min} = \min\Delta\phi(\met,jet)$,
defined as the minimum angular difference between the $\met$ and each 
jet in the event. This quantity is expected to be large 
in $W\to l\nu$ decays and 
in $t\bar t\to \met$ + jets events. 
On the other hand, for QCD background events where the main source 
of $\met$ is due to jet energy mismeasurement, the $\met$ is expected to be 
aligned with the jet direction and the value of
$\Delta\phi_{min}$ close to zero.

Other kinematic variables related to the topology 
of the event characterize the $t\bar t$ production 
with respect to background processes.
Let $Q_{j(j=1,3)}$ be the eigenvalues of the 
normalized momentum tensor
$M_{ab} = \frac{\sum_j P_{ja} P_{jb}}{\sum_j P_j^2}$~\cite{ColliderPhysics},
where $a,b$ run over the three space coordinates, and $P_j$ is the momentum
of the jet~$j$. The sphericity $S$ is defined as 
$S = \frac{3}{2} (1 - Q_3)$. $S$ is zero in the limiting case
of a pair of back-to-back jets,
while approaches one for events with a perfectly isotropic 
jet momenta distribution. 
The aplanarity $A$ is defined as
$A = \frac{3}{2} Q_1$ and lies in the range $[0,\frac{1}{2}]$.
Extremal values of $A$ are 
reached in the case of two opposite jets and in the case of evenly
distributed jets, respectively. 
Jets emerging from a $t\bar t$ pair are expected to be uniformly distributed
and, as a consequence, they will hardly lie on the same plane: thus we
expect high aplanarity and sphericity values for $t\bar t$ events.
In addition to kinematic variables describing the topology 
of the event,
distributions of energy-related variables
%,such as the centrality, $\sum E_T$ and $\sum_{3} E_T$, 
can be useful
to discriminate $t\bar t$ events over 
their background. 
The centrality $C$ is defined as 
$C = \frac{\sum E_T}{\sqrt{\hat{s}}}$, where 
$\sqrt{\hat s}$ is the invariant mass of the jets system.
In the case of $t\bar t$ pairs decaying hadronically, 
jets are emitted preferably in the transverse plane ($r-\phi$ plane), so
we expect to have a greater amount of energy emitted in this plane
giving values of $C$ closer to 1 with respect to background events.
The variable $\sum_{3} E_T$ is defined as the sum of all 
jets $E_T$ in the event except the two 
leading ones. In QCD events, the two most energetic jets
are produced by $q\bar q$ processes, while the least energetic ones
come from gluon \emph{bremsstrahlung}; on the contrary,
in $t\bar t$ events up to six jets can be produced by hard processes,
and as a consequence $\sum_{3} E_T$ can help discriminating
signal and background. 
%Another kinematic variable we can exploit is $E_{T}^{1}$, the transverse 
%energy of the leading jet in the event.
All these variables will be used in the following section
to train a neural network to discriminate $t\bar t$
signal from background.

\section{Neural network based event selection}\label{sec:SignalSel}
In order to enhance the signal to 
background ratio in the data sample, we use a neural
network (NN), trained to discriminate $t\bar t \to \met$~+~jets signal
events from background. The NN is built using the  
neural network implementation in {\scshape{root}}~\cite{bib_root}.

\begin{figure*}%[Ht!]
\begin{center}
 \subfigure[][]{\label{subfig:et1} \includegraphics[width=.3\linewidth]{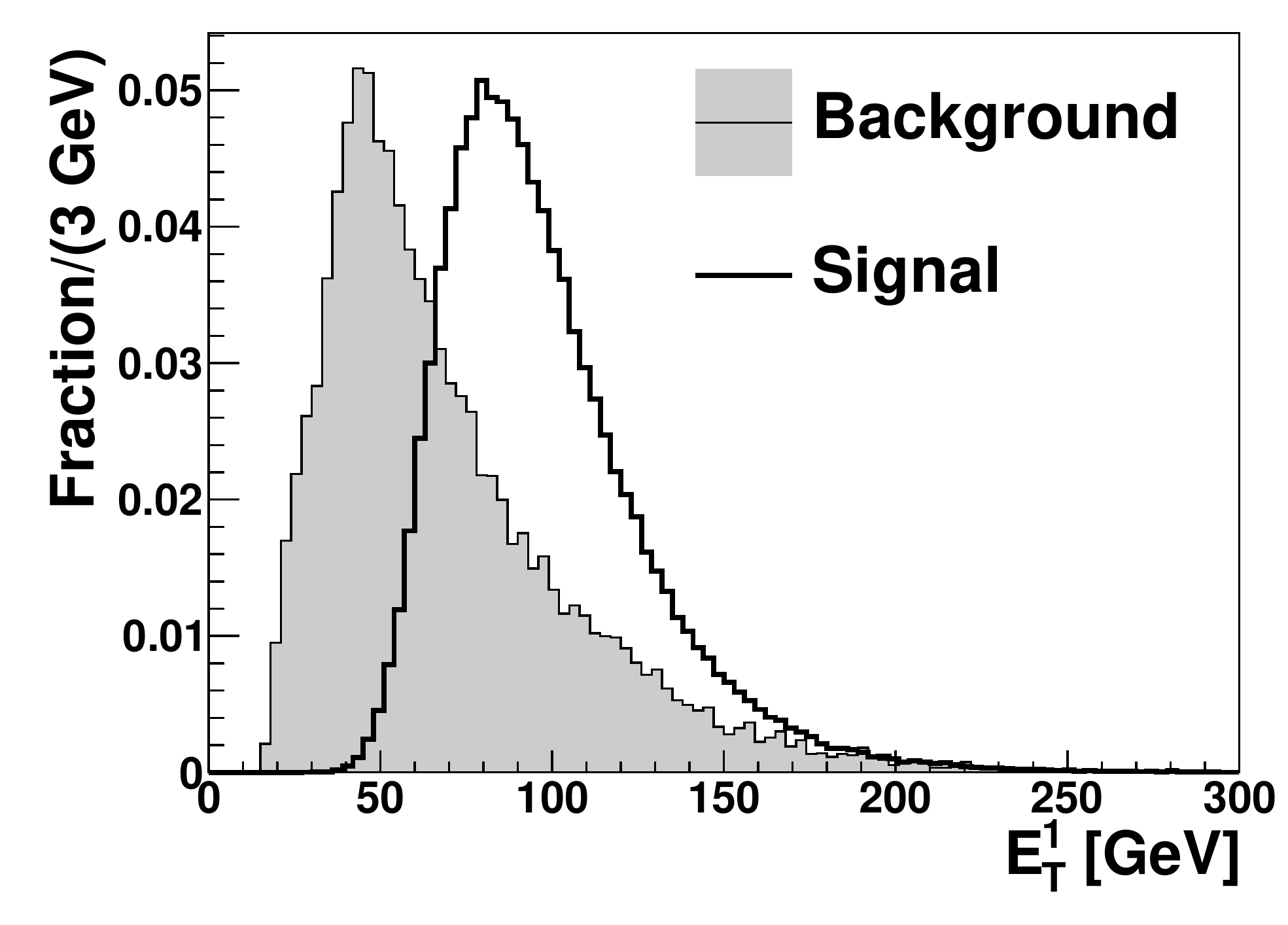}}
 \subfigure[][]{ \label{subfig:dphimin}\includegraphics[width=.3\linewidth]{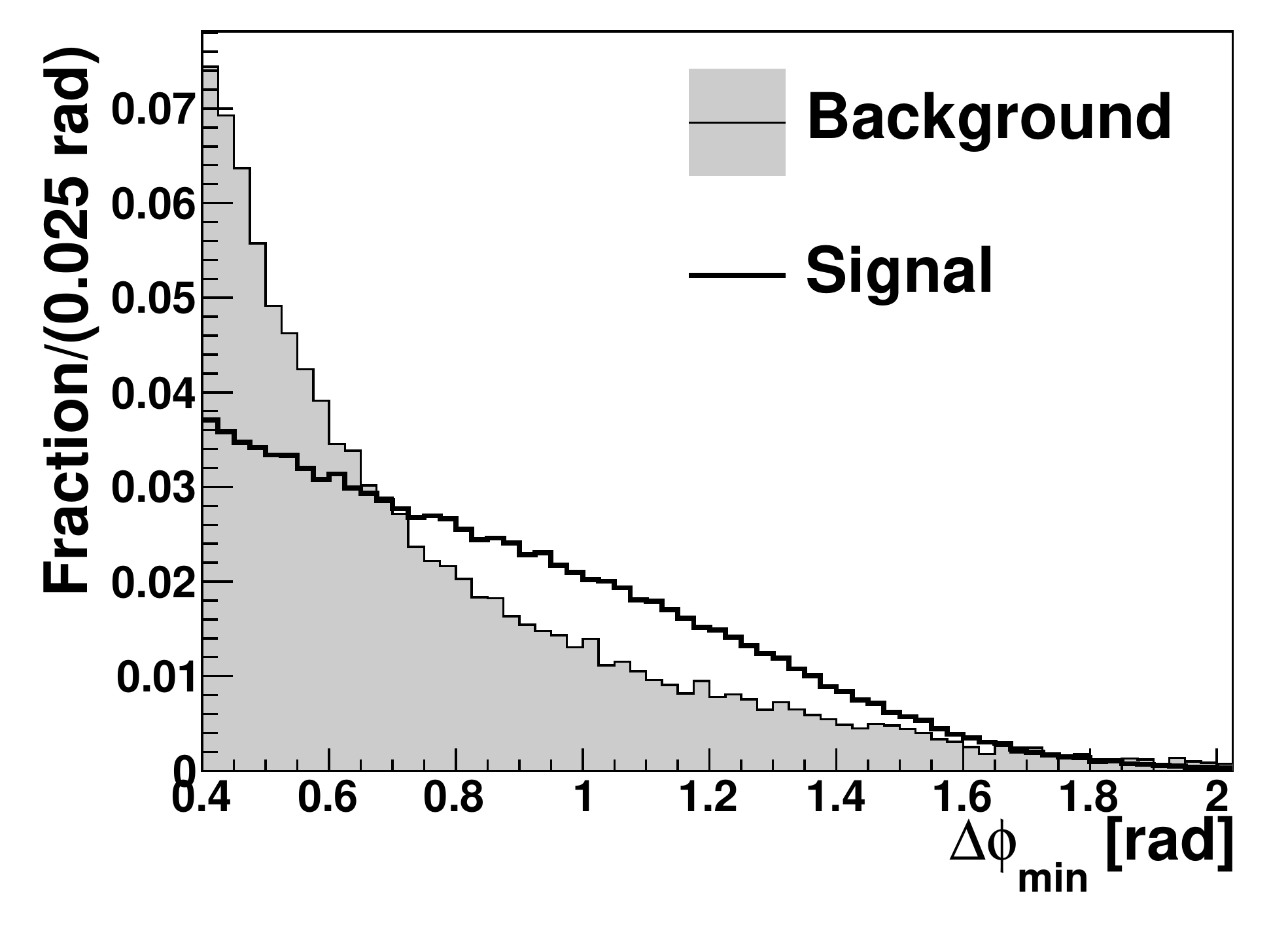}}
 \subfigure[][]{ \label{subfig:mets}\includegraphics[width=.3\linewidth]{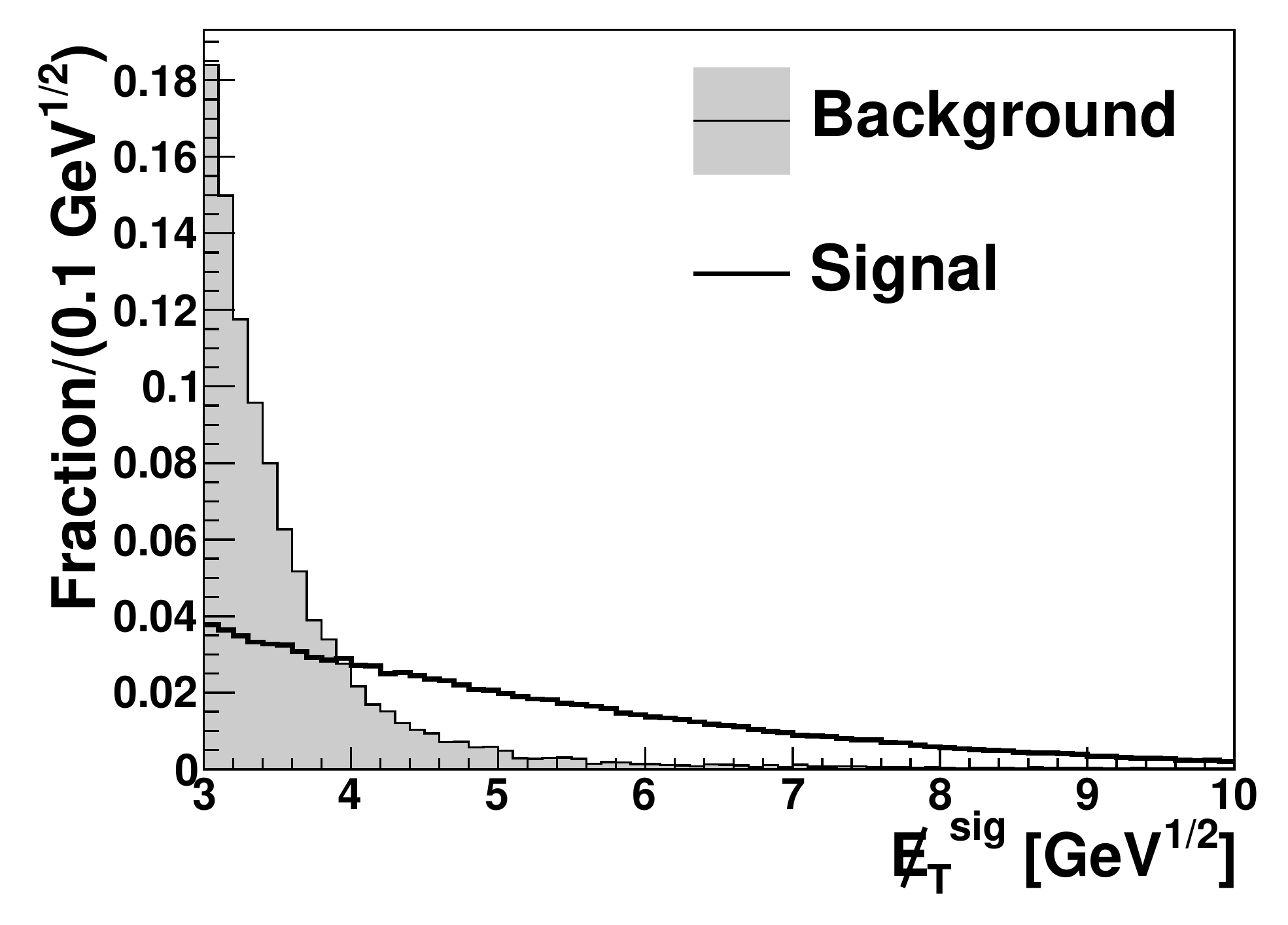}}
 \subfigure[][]{ \label{subfig:sumet}\includegraphics[width=.3\linewidth]{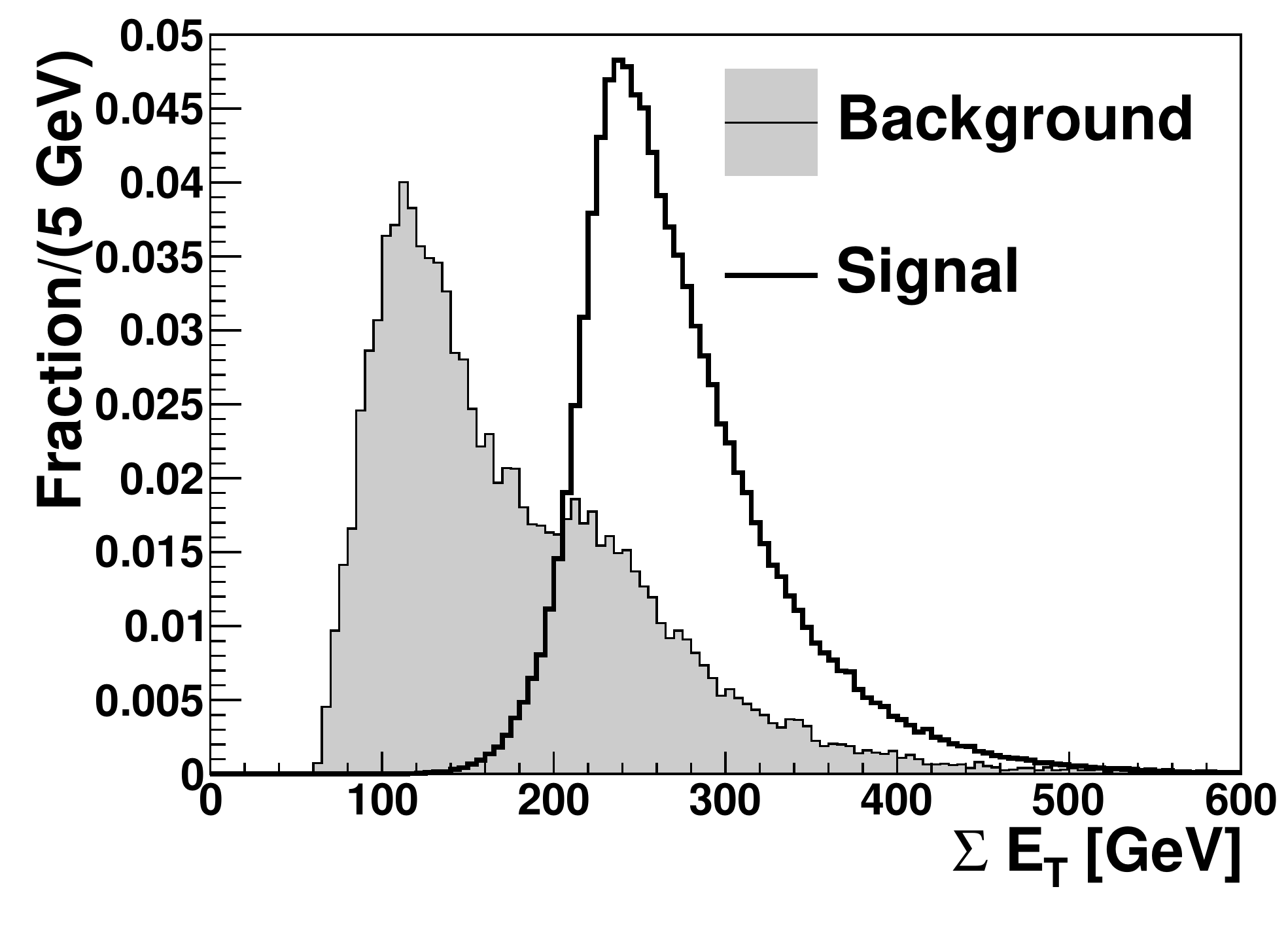}}
 \subfigure[][]{ \label{subfig:sumet3}\includegraphics[width=.3\linewidth]{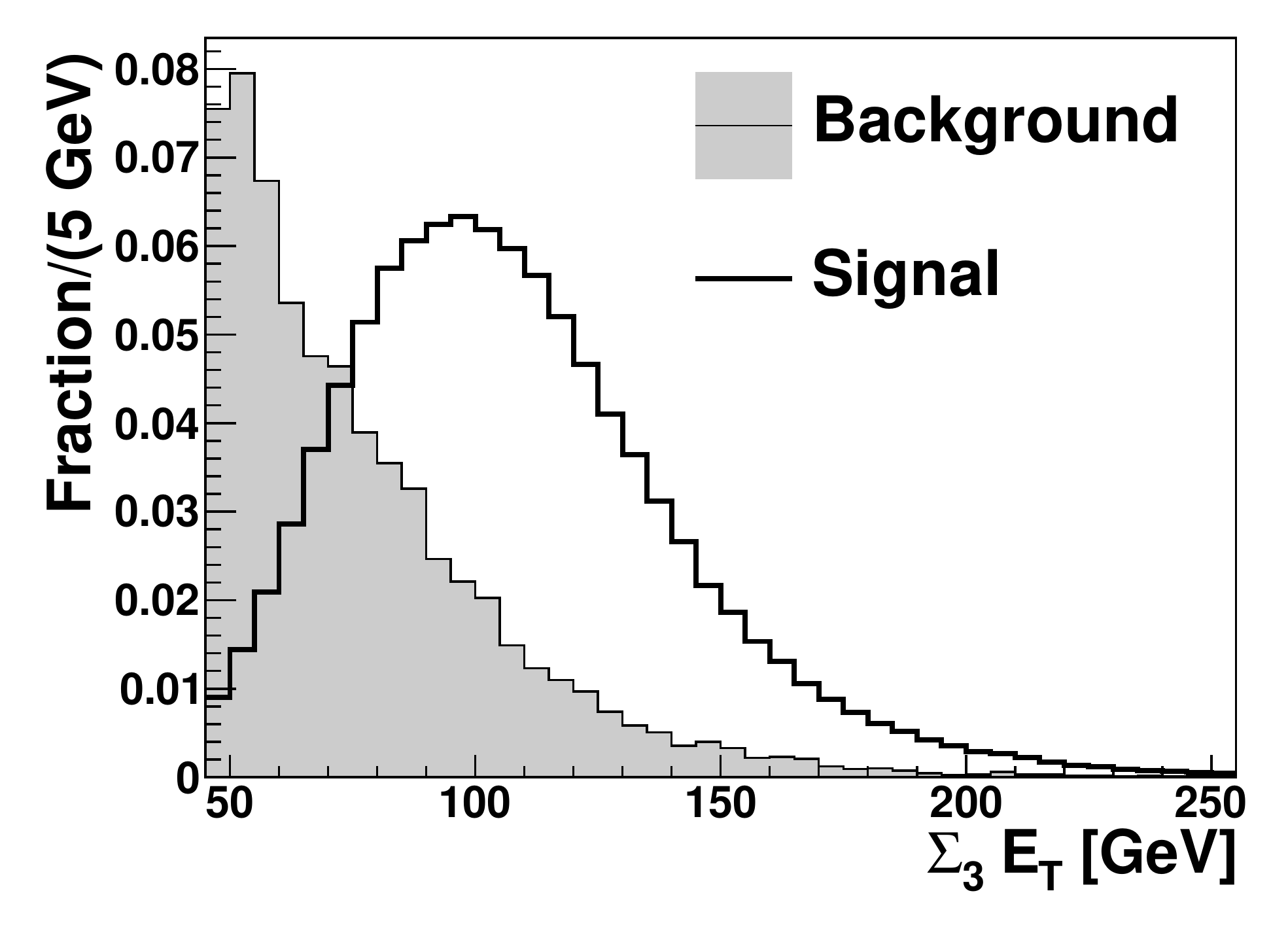}}
 \subfigure[][]{ \label{subfig:cent}\includegraphics[width=.3\linewidth]{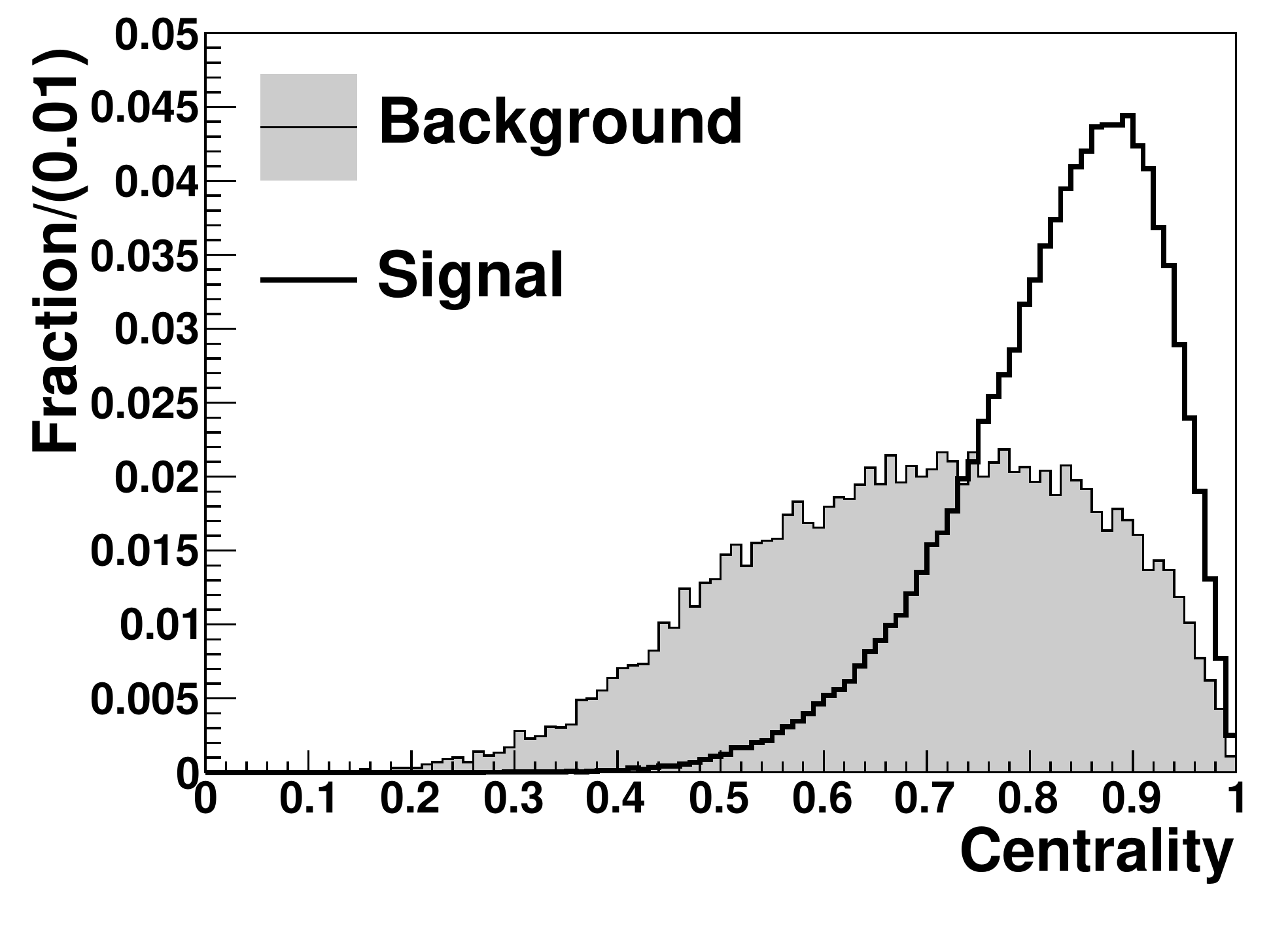}}
 \subfigure[][]{ \label{subfig:spher}\includegraphics[width=.3\linewidth]{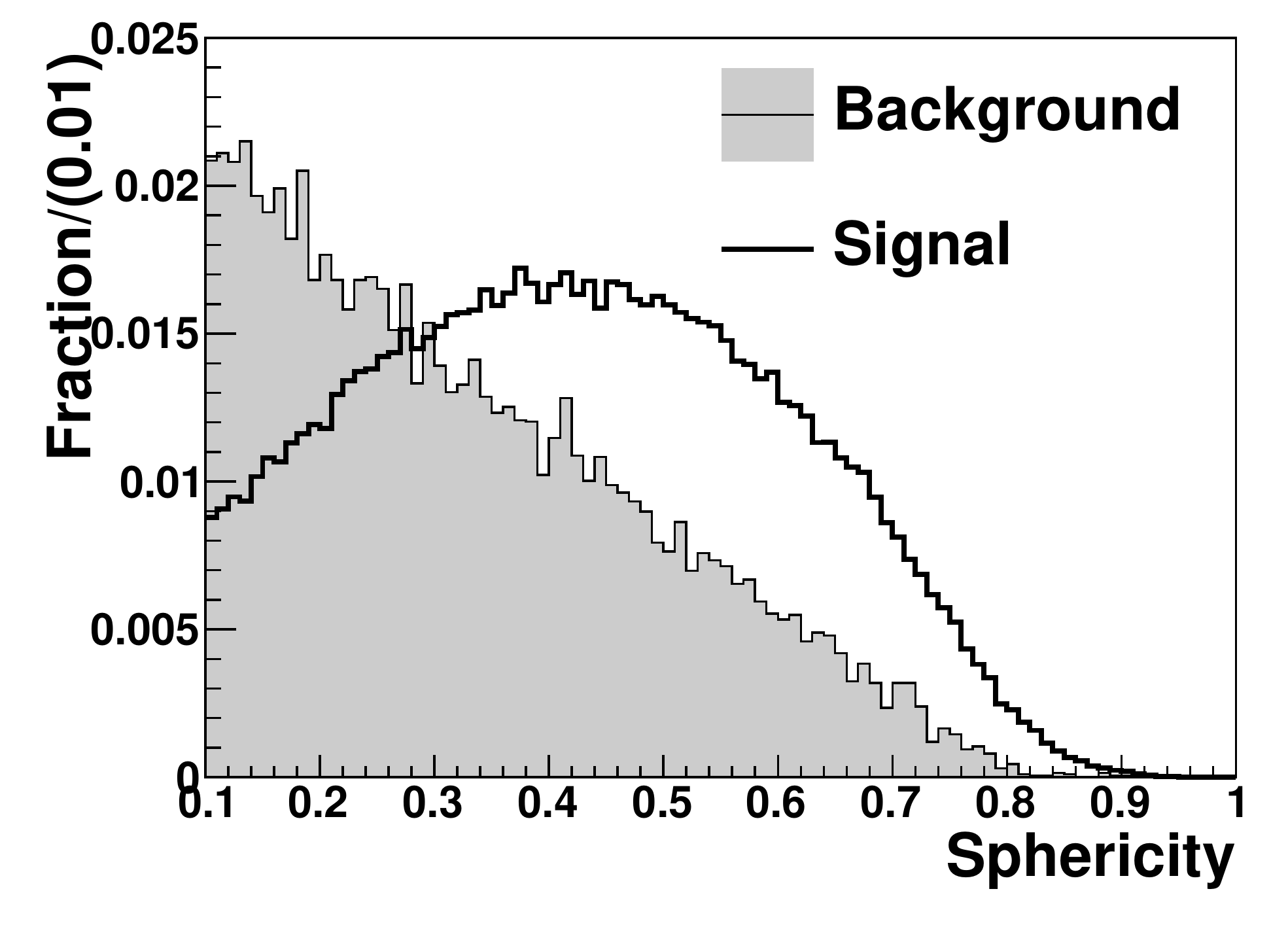}}
 \subfigure[][]{ \label{subfig:apla}\includegraphics[width=.3\linewidth]{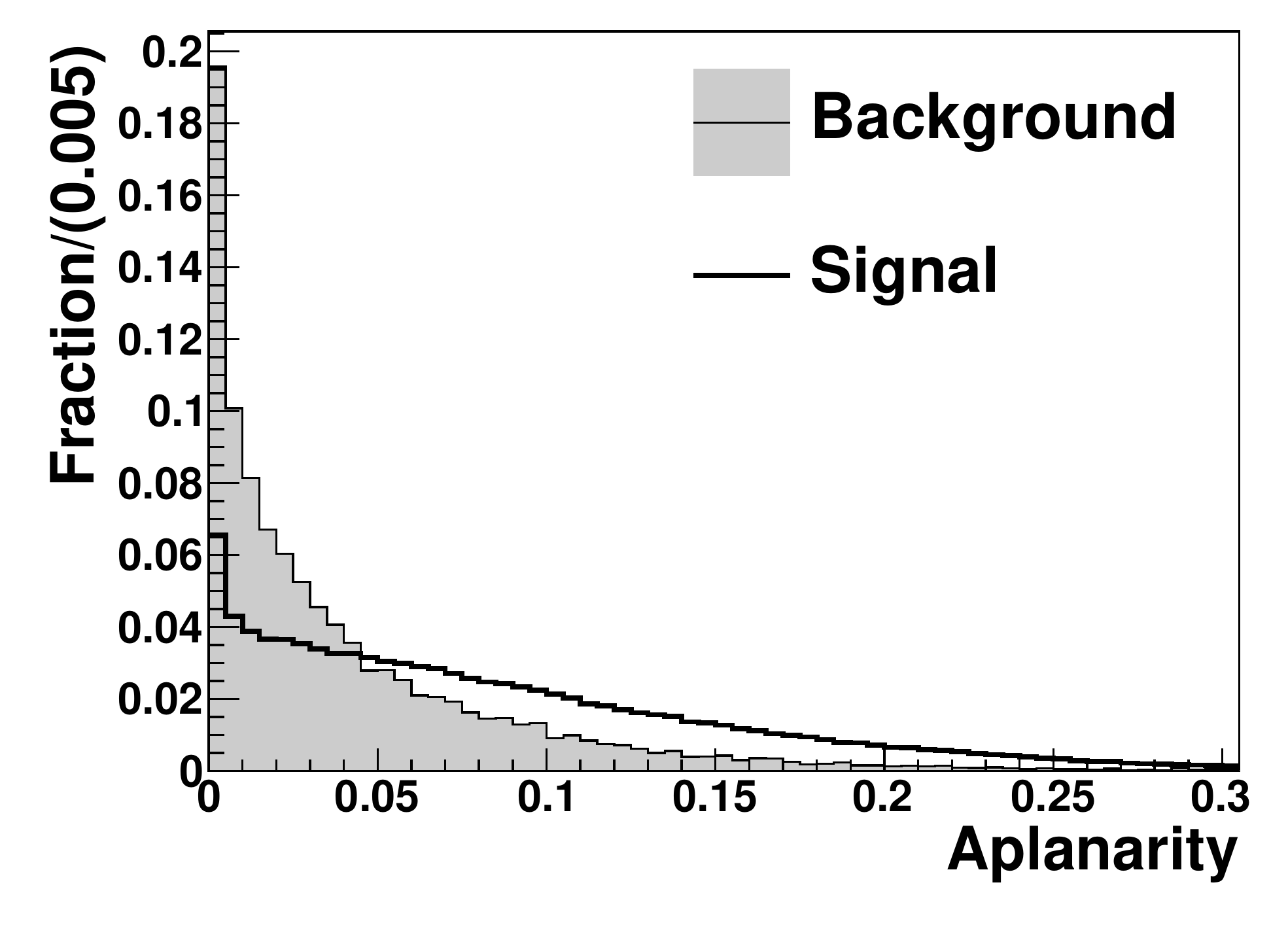}}
 \end{center}
 \caption{Distribution of neural network input variables for 
 	multijet data (background) and 
	$t\bar t$ MC (signal) samples:  the transverse energy of the 
  leading jet $E_{T}^{1}$~\subref{subfig:et1}, 
  $\Delta\phi_{min}(\met,jet)$~\subref{subfig:dphimin},  
  ${\met}^{sig}$~\subref{subfig:mets}, 
  $\sum E_T$~\subref{subfig:sumet},
  $\sum_{3} E_T$~\subref{subfig:sumet3},
  the centrality $C$~\subref{subfig:cent},
 and the topology-related variables sphericity $S$~\subref{subfig:spher}
 and aplanarity $A$~\subref{subfig:apla}.}
 \label{fig:nn_input}
 \end{figure*}

We apply an additional kinematic requirement
on data with at least four jets,
removing events with low angle between 
jets and ${\met}$ with the cut 
$\Delta\phi_{min} (\met, jets)>0.4$.
In the data these events come mostly from mismeasured 
jets and are difficult to simulate in MC. 
They are removed from the NN training to prevent it from converging 
on artificial differences between MC and data 
rather than real physics effects. 
After this requirement, we are left with
$20~043$ events in the  sample with at least four jets,
with an expected $S/B$ of $3.5\%$:
since this sample has low $t\bar t$ signal fraction
and no overlap with data used to 
determine the background parametrization,
it will be used as background training sample.
As signal training sample we use the same amount of events 
passing the same event selection of data, taken randomly
from the $t\bar t$ simulation.
%as this allows the best training performances of the class used.
We use as inputs for the network the following kinematic variables, 
normalized with respect to their maximum values:
 the transverse energy of the leading jet $E_{T}^{1}$,
$\Delta\phi_{min}(\met,jet)$, 
${\met}^{sig}$, 
$\sum E_T$, $\sum_{3} E_T$,
the centrality $C$,
and the topology-related variables sphericity $S$ and aplanarity $A$.
The distributions of the input variables used in the NN
for both the signal and background training samples are shown in Fig.~\ref{fig:nn_input}.

\begin{figure*}%[htbp]
\center
\subfigure[][]{\label{subfig:al4jets}\includegraphics[width=0.45\linewidth]{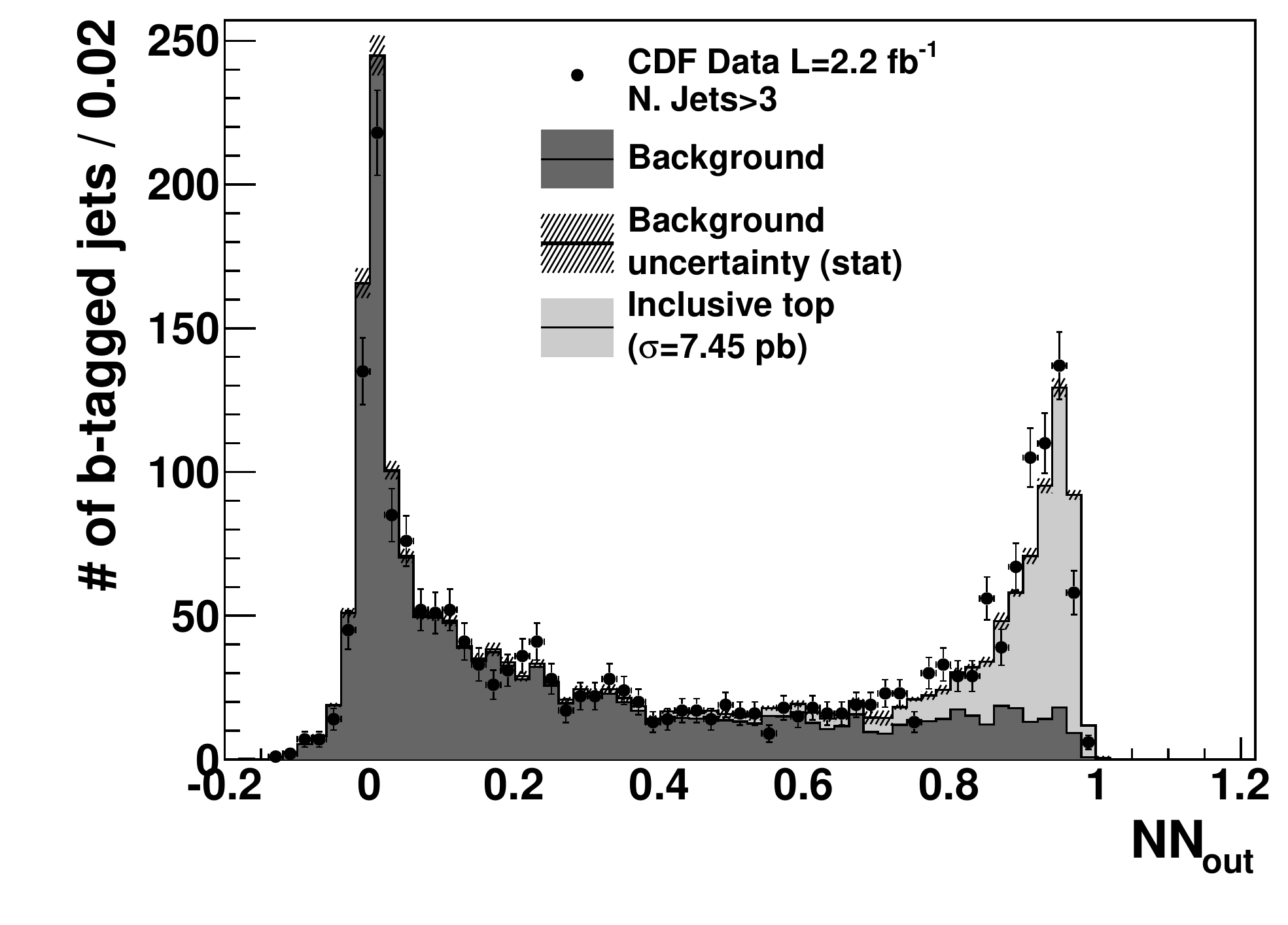}}
\subfigure[][]{\label{subfig:3jets}\includegraphics[width=0.45\linewidth]{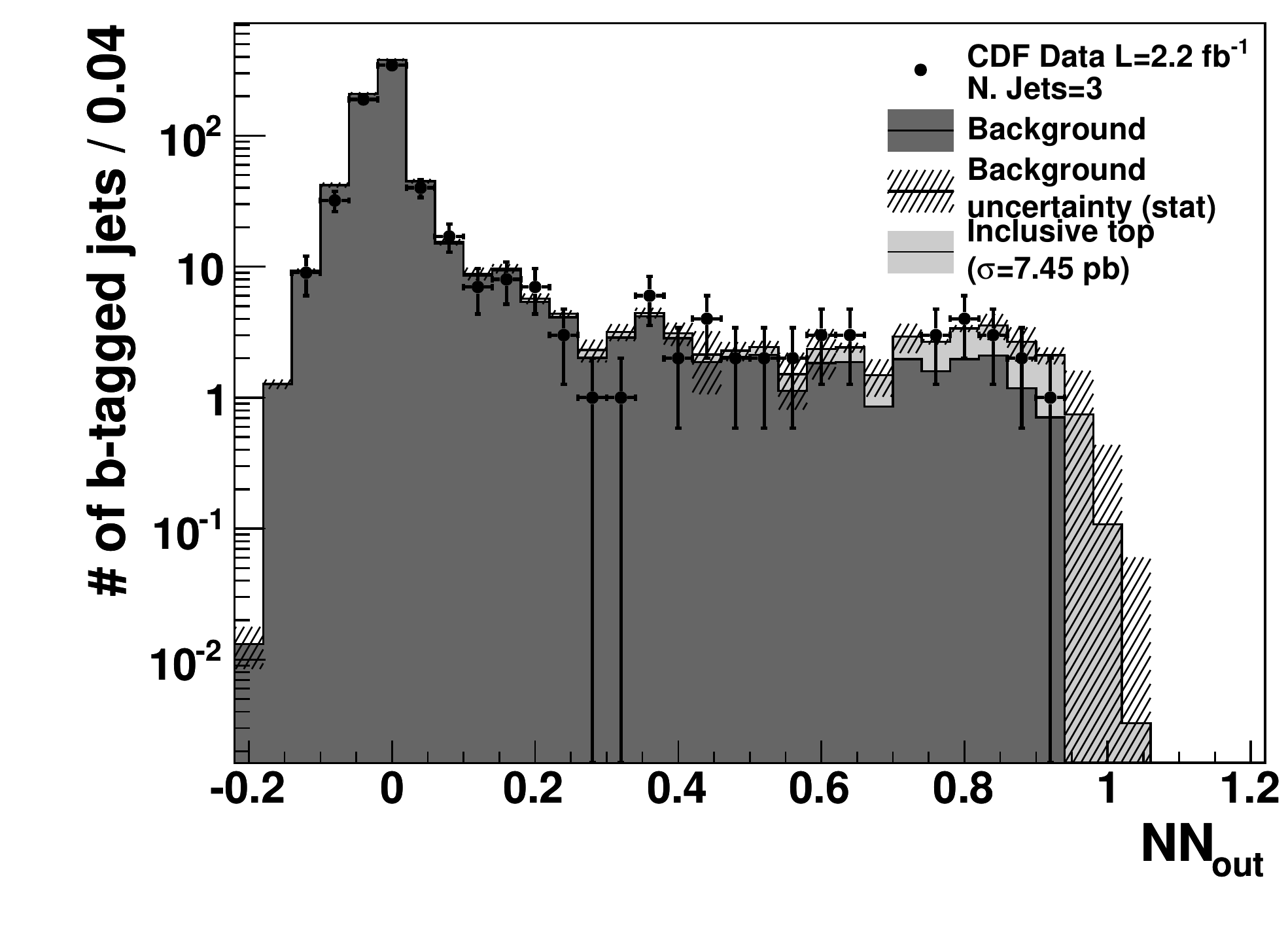}}
\subfigure[][]{\label{subfig:4jets}\includegraphics[width=0.45\linewidth]{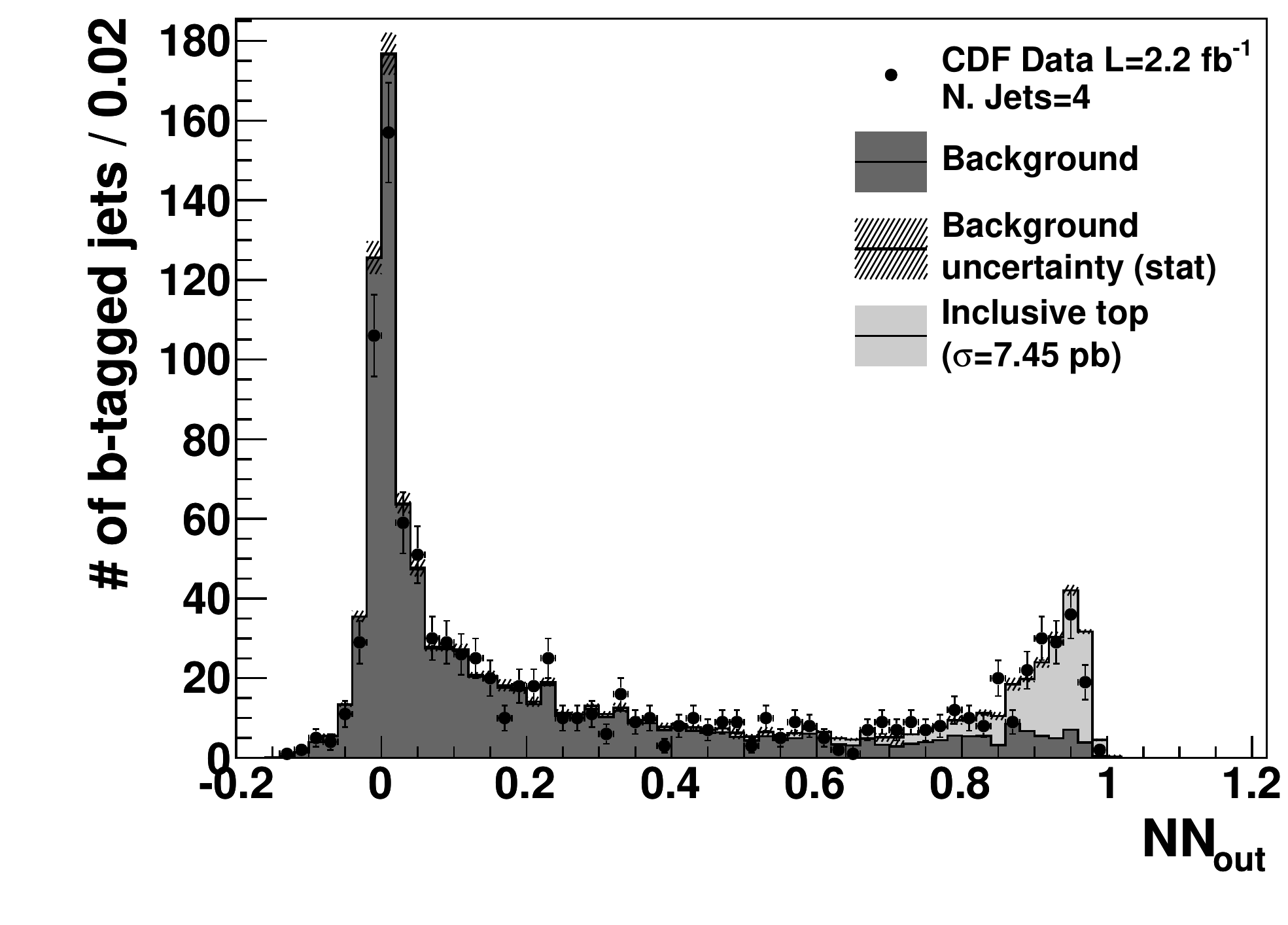}}
\subfigure[][]{\label{subfig:5jets}\includegraphics[width=0.45\linewidth]{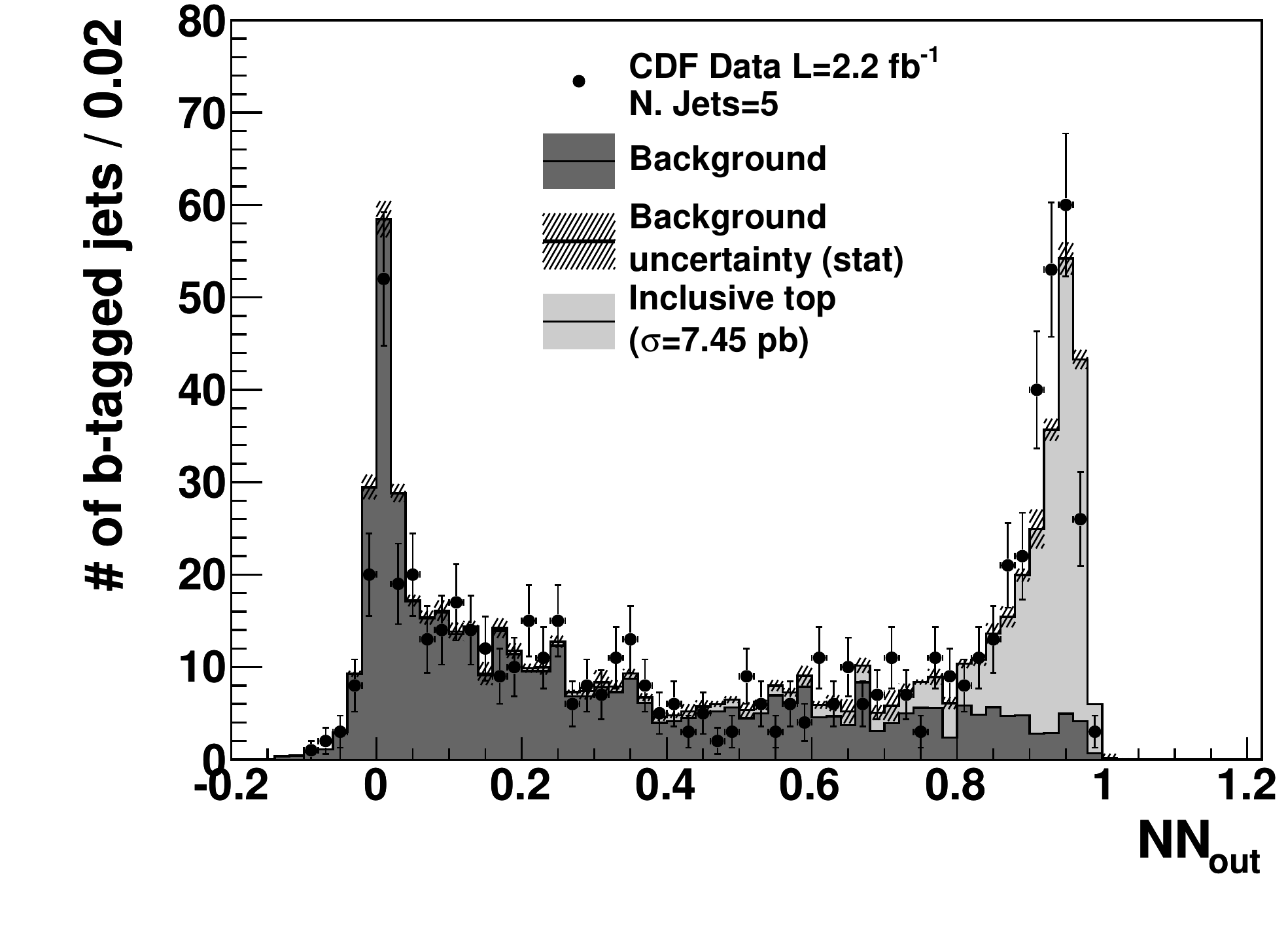}}
\caption{Observed and matrix-predicted number of $b$-tagged
	jets versus neural network output in the multijet data for all events
	with at least four jets~\subref{subfig:al4jets} and for events with exactly
	three~\subref{subfig:3jets}, four~\subref{subfig:4jets} 
	and five~\subref{subfig:5jets} jets, 
    along with the expected contribution 
	due to $t\bar t$ signal events.
	Background error bands are centered on the 
    inclusive top plus background prediction.} 
\label{fig:tags_nnout}
\end{figure*}

After the training, the $\bq$-tagged data are processed by the NN:
Fig.~\ref{fig:tags_nnout} shows the  number of observed $b$-tagged 
jets versus the NN
output NN$_{out}$,
along with the corresponding background prediction 
from the tag rate parametrization
and expected contribution from $t\bar t$ signal ($m_{top} = 172.5$~GeV/$c^2$), 
for events with at least four jets
and with exactly three, four and five jets.
The good agreement between data and
the sum of expected background and $t\bar t$ induced $b$-tagged jets
in the high NN output region
is both a confirmation of the effectiveness of the
method we use to estimate the background and 
a demonstration of proper NN training and performance.

%%%%%%%%%%%%%%%%%%
%%%%%%%%%%%%%%%%%%

In order to select a signal-rich sample
to perform the cross section measurement,
we choose to cut on the NN output value NN$_{out}$.
The cut is chosen
with the aim of minimizing 
the statistical uncertainty on the cross section measurement,
maximizing $S/B$ where  
$S$ is the expected number of $b$-tagged jets from the signal
and $B$ is the predicted number of background $b$-tagged jets.
The former quantity is evaluated from an
inclusive $t\bar t$ MC sample, while the latter is derived 
using the $b$-tagging matrix parametrization on data.  
The result of this optimization procedure 
is a cut on 
NN$_{out} \ge 0.8$, which gives the lowest expected statistical
uncertainty on the cross section ($\sim 8\%$) and an expected $S/B \sim 4$.
The expected signal sample composition
after this cut is shown in Table~\ref{tab:sample_composition}.

\begin{table*}
\caption{Expected signal sample composition after the NN$_{out} > $ 0.8 cut.}
 \begin{center}
 \begin{tabular}{c c c c c c c c r}
  \hline\hline
  Number of jets & 3 & 4 & 5 & 6 & 7 & 8 & 9 & Total\\
  \hline
all-hadronic (\%) & 0.1  & 0.5  & 1.7  & 4.9  & 7.8  & 10.2  & 9.9  & 2.3  \\
$e$ +jets (\%) & 26.7 & 25.2  & 35.5  & 36.2  & 35.1  & 33.6  &
33.8  & 32.1  \\
$\mu$ +jets (\%) & 32.2  & 32.5  & 19.1  & 15.8  & 14.5  & 16.1
& 12.9  & 22.7  \\
dileptonic (\%) & 6.5  & 2.3  & 1.1  & 0.7  & 0.5  & 0.6  & 0.00  & 1.4  \\
hadronic $\tau$ +jets (\%) & 15.8  & 21.9  & 30.5  & 31.5  & 31.8  &
29.9  & 34.4  & 27.7 \\
leptonic $\tau$ +jets (\%) & 11.7  & 13.2  & 9.9  & 9.1  & 8.9  & 8.1
 & 8.4  & 10.8  \\
$\tau\tau$ (\%) & 1.1  & 1.1  & 0.6  & 0.5  & 0.4  & 0.4  & 0.3
 & 0.8  \\
$e$/$\mu$ + $\tau$ (\%) & 6.0  & 3.4  & 1.6  & 1.3  & 0.9 & 1.1
 & 0.3  & 2.2  \\
  \hline\hline
  \end{tabular}\label{tab:sample_composition}
 \end{center}
\end{table*}

\section{Systematic uncertainties}\label{sec:Syst}
The top quark pair production cross section
is measured as:
\begin{equation}
\sigma(p\bar p \to t\bar t) \times BR(t\bar t \to \met+jets) = 
   \frac{N_{obs} - N_{exp}}{ \epsilon_{kin} \cdot 
    \epsilon_{tag}^{ave} \cdot L} 
\label{eq:xseccval}
\end{equation}
where $N_{obs}$ and $ N_{exp}$ are the number of observed and 
predicted tagged jets from background in the selected sample, respectively; 
$\epsilon_{kin}$ is the kinematic efficiency of trigger, preliminary 
requirements and neural network selection  
determined using inclusive MC $t\bar t$ events; 
$\epsilon_{tag}^{ave}$,~defined 
as the ratio of the number of $b$-tagged jets to the number
of events in the inclusive $t\bar t$ Monte 
Carlo sample, gives the average number of $b$-tagged jets per $t\bar t$ event. 
Finally, $L$ is the integrated luminosity of the dataset used.
All quantities in the denominator of Eq.~\ref{eq:xseccval}, as
well as the number of expected  $b$-tagged jets, are 
subject to different sources of systematic uncertainties.

The kinematic efficiency $\epsilon_{kin}$ is 
evaluated on inclusive $t \bar t$ samples generated with 
{\sc{pythia}} and
its uncertainty arises from the particular 
choice of the MC generator,
the set of parton density functions used in the generator, 
as well as the modeling of color reconnection effects and of the
initial and final state radiation.

The MC generators differ in their hadronization 
schemes and in their description of the underlying event and multiple interactions:
in order to evaluate the generator dependence of 
the kinematic efficiency, we compare the value of $\epsilon_{kin}$ obtained 
using {\sc{pythia}}  with the value obtained on a sample of events
generated with {\sc{herwig}} ~v6.510~\cite{herwig}, and take
the relative difference as the systematic uncertainty. \\
The choice of parton distribution function~(PDF) 
affects the kinematics of 
$t\bar t$ events, and thus the acceptance for signal events.
We estimate this uncertainty 
comparing the $\epsilon_{kin}$ value derived
from MC samples based on the default 
PDF~CTEQ5L~\cite{CTEQ5L} 
with the one obtained using samples based on
MRST72 and MRST75~\cite{MRST}, which differ by the value of  
the strong coupling constant $\alpha_s$ used to compute the PDF;
we also consider the difference
in the value of $\epsilon_{kin}$ obtained with the
leading order~(LO) and next to leading order~(NLO) calculations
of PDFs, evaluated using
default CTEQ5L~(LO) and CTEQ6M (NLO) PDFs, and
derive the corresponding uncertainty.
We add these two contributions in quadrature to obtain
the systematic uncertainty due to our choice of PDF.\\
Uncertainties arising from the modeling of color reconnections effects
are estimated by evaluating the shift in the kinematic efficiency 
using two  samples of events generated by {\sc{pythia}},
corresponding to different 
models of color reconnection~\cite{color-reconnection}.\\
Additional jets coming from initial and final state radiation (ISR and FSR), 
might change the sample composition and affect the
efficiency of the kinematic selection.
The systematic uncertainty related to these effects 
is evaluated by calculating the shift in $\epsilon_{kin}$
using inclusive 
$t\bar t$ samples
with different amounts of  
initial and final state radiation.\\
The systematic uncertainty due to the calorimeter response
is accounted for by varying the corrected jet energies 
within $\pm 1\sigma$ of their corresponding systematic uncertainty, and 
recalculating $\epsilon_{kin}$ after these variations.\\
Finally, $\epsilon_{kin}$ is also affected by the simulation
of the trigger requirements on MC events and a 
trigger acceptance uncertainty 
is determined by 
comparing trigger turn-on curves between MC
and data events.

The average number of $b$-tagged jets 
per $t\bar t$ event is affected by the uncertainty on the scale factor
used to account for the different 
efficiency of the $b$-tagging algorithm in data
and in MC. 
The systematic uncertainty on  $\epsilon_{tag}^{ave}$
is obtained varying the scale factor 
within the $\pm 1 \sigma$ range from its central value
and determining,  on the MC sample, the difference in terms 
of average number of $b$-tagged jets per event with respect 
to the standard $\epsilon_{tag}^{ave}$ value.\\
The uncertainty on the number $N_{exp}$ of matrix-predicted $b$-tagged 
jets is calculated by comparing the number of $b$-tagged jets yielded by the tagging
matrix to the actual number  
of observed  {\scshape{secvtx}} tagged jets in a control sample depleted of signal 
(events with NN output lower than 0.6). 
The relative difference between the expected and observed 
number of tagged jets
is taken as an estimate of the uncertainty of our background prediction.\\
The luminosity measurement is affected by two sources of 
uncertainty: 
the acceptance of the luminosity monitor  
and the total inelastic $p\bar p$ cross section ($60.7\pm 2.4$~mb).
The uncertainties on these quantities are $4.2\%$ 
and $4.0\%$ respectively, giving a total uncertainty of $5.8\%$ on the 
integrated luminosity calculated for any given CDF dataset. \\
The summary of all sources of systematic uncertainties to the 
cross section evaluation is listed in Table~\ref{tab:syst_tot}.

\begin{table}
\caption{Summary of relative systematic uncertainties on the signal 
efficiency, and other uncertainties related to the cross section evaluation.}
\begin{center}
\begin{tabular}{l r}
\hline\hline
Source & Uncertainty \\
\hline
\multicolumn{2}{c}{$\epsilon_{kin}$ systematics}\\
Generator dependence & 3.9 \% \\
PDFs                 & 1.2 \% \\
ISR/FSR              & 2.7 \% \\
Color Reconnection   & 4.3 \% \\
Jet Energy Scale     & 4.2 \%  \\
Trigger simulation   & 3.0 \% \\
\multicolumn{2}{c}{Other systematics} \\
$\epsilon_{tag}^{ave}$ ($\bq$-tag scale factor) & 3.9 \%   \\
Background parametrization & 2.5 \% \\
Luminosity measurement      & 5.8 \%    \\
\hline\hline
\end{tabular}\label{tab:syst_tot}
\end{center}
\end{table}

\begin{table}[Ht!]
\begin{center}
\caption{Input values for the cross section measurement. $\epsilon_{tag}^{ave}$ is   the average number of $b$-tagged jets per top event.}
\begin{tabular}{l l c}
\hline\hline
Variable & Symbol & Value  \\
\hline
Integrated Luminosity ($pb^{-1})$ & $L$     & $2207.5 \pm 128$ \\
Observed $\bq$-tagged jets             & $N_{obs}$         & $636$ \\
Background $\bq$-tagged jets           & $N_{bkg}$  &
$131 \pm 9.6$   \\
Kinematic efficiency (\%)              & $\epsilon_{kin}$ & $3.53 \pm 0.29$   \\
$b$-tagged jets per $t\bar t$ event (\%)      & $\epsilon_{tag}^{ave}$  & $81.1
\pm 3.2$ \\
\hline\hline
\end{tabular}\label{tab:measurement}
\end{center}
\end{table}

\section{Cross section measurement}\label{sec:XsecMeas}
After the neural network 
selection we are left with $1420$ events with at least four jets 
of which  $636$ are $b$-tagged.
Inserting in  Eq.~\ref{eq:xseccval} the input parameters quoted in 
Table~\ref{tab:measurement}, the measured cross section value is

\begin{eqnarray}
\sigma _{t\bar t} &=&  7.99~\pm 0.55 \mbox{ (stat) } \pm 0.76  \mbox{
(syst) } \pm 0.46  \mbox{ (lumi)}~\mbox{pb} \nonumber\\
                 &=& 7.99~\pm 1.05~\mbox{pb} \nonumber
\end{eqnarray}

\begin{figure}%[htbp]
\begin{center}
\includegraphics[width=0.9\linewidth]{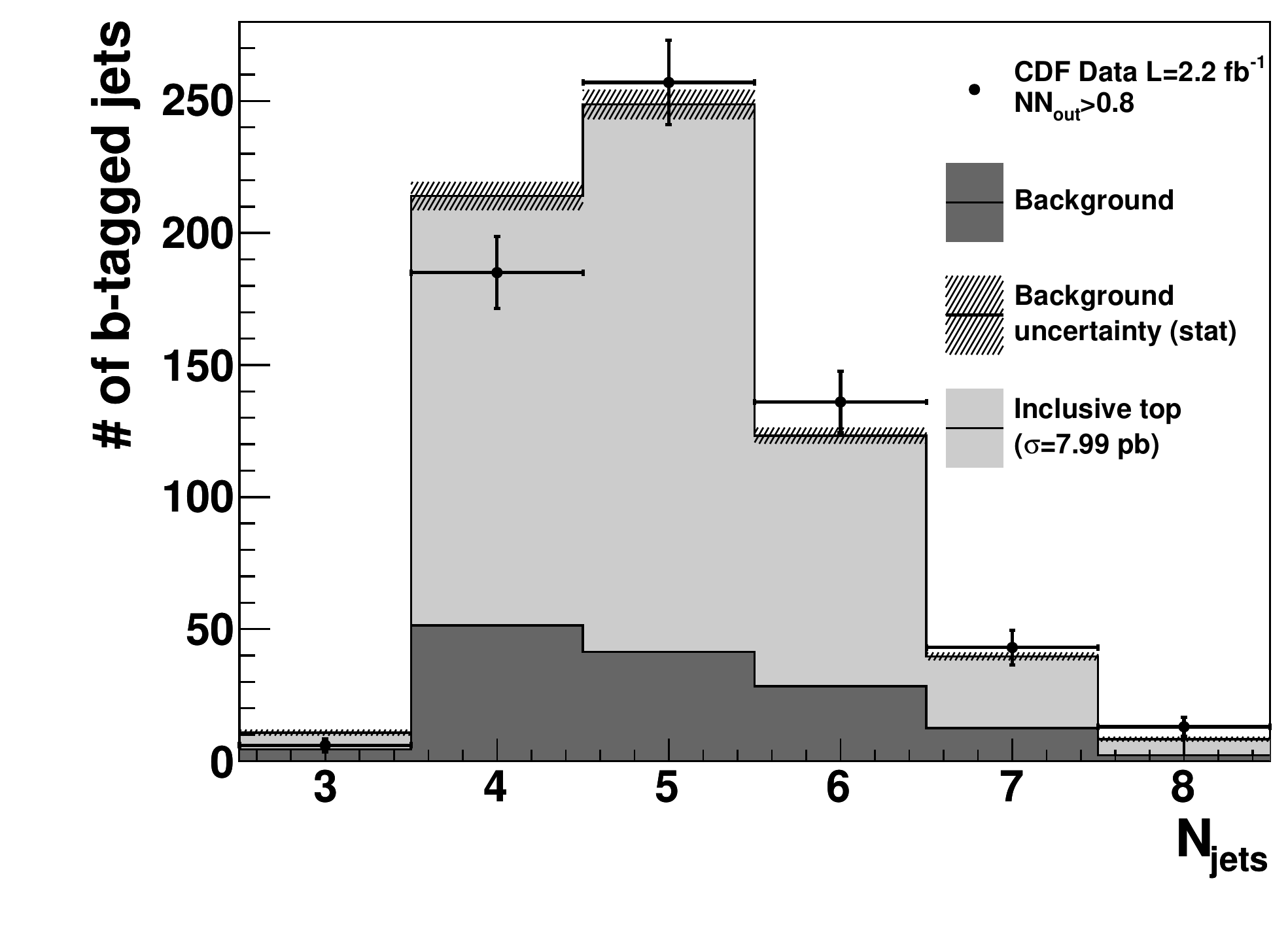}
\caption{Observed and matrix-predicted number of $b$-tagged
jets by jet multiplicity in the multijet data after cut on neural 
network output NN$_{out}$ greater than 0.8,
along with the expected contribution of inclusive $t\bar t$ signal
normalized to the measured cross section. 
Background error bands are centered on the 
inclusive top plus background prediction.} 
\label{fig:tags_after_nncut}
\end{center}
\end{figure}

Observed and expected $b$-tagged jets after selection for different
jet multiplicities are shown in Fig.~\ref{fig:tags_after_nncut}, along
with the expected contribution of inclusive $t\bar t$ signal,
normalized to the measured cross section.

%%%%%%%%%%%%%%%%%%
%%%%%%%%%%%%%%%%%%

\section{Conclusions}
We presented a measurement of the $t \bar t$ production cross section in 
 a final state with large missing transverse energy and multiple jets.
 We explicitly vetoed well identified high-$p_T$ electrons 
or muons from $W$ boson decay, and rejected events with low missing
 transverse energy, to avoid overlaps with other cross section 
measurements performed  by the CDF Collaboration.
Using an optimized neural network based kinematic selection
and a $\bq$ jet identification technique
on a sample of data corresponding to an integrated luminosity of 2.2 fb$^{-1}$, 
we obtain a production cross section 
value of $7.99\pm1.05$~pb, in good agreement with 
 the reference theoretical value of $7.46^{+0.66}_{-0.80}$~pb~\cite{top_th}
 and with other next to next to leading order calculations~\cite{other-th} 
for a top mass of $172.5$~GeV/$c^2$. The agreement with the most recent
experimental determinations is also good. Given the high precision of 
the result, this independent cross section measurement will have a significant
impact on a future CDF combined value. 
The same $\met$ + jets selection
can be used for top quark mass measurement
to obtain a result statistically uncorrelated 
with those of other methods.

\begin{acknowledgments}
We thank the Fermilab staff and the technical staffs
of the participating institutions for their vital contributions.
This work was supported by the U.S. Department of Energy and National
Science Foundation; the Italian Istituto Nazionale di Fisica Nucleare;
the Ministry of Education, Culture, Sports, Science and Technology
of Japan; the Natural Sciences and Engineering Research Council
of Canada; the National Science Council of the Republic of China;
the Swiss National Science Foundation; the A.P. Sloan Foundation;
the Bundesministerium fuer Bildung und Forschung, Germany;
the Korean Science and Engineering Foundation and the Korean
Research Foundation; the Particle Physics and Astronomy Research
Council and the Royal Society, UK; the Russian Foundation for
Basic Research; the Comision Interministerial de Ciencia y
Tecnologia, Spain; and in part by the European Community's
Human Potential Programme under contract HPRN-CT-2002-00292,
Probe for New Physics.

\end{acknowledgments}

\end{document}